\documentclass[journal,web,twoside]{ieeecolor}

\usepackage{generic}
\usepackage{algorithmic}
\usepackage{graphics}
\usepackage{arydshln}
\usepackage{textcomp}

\usepackage{amsmath,amssymb,amsthm,xpatch,amsfonts,dsfont}
\usepackage{cite}
\usepackage{graphicx}
\usepackage{textcomp}
\usepackage{epstopdf}
\usepackage[font=footnotesize]{caption}
\usepackage{psfrag}
\usepackage{subfigure}
\usepackage{multirow}
\usepackage{float}

\newtheorem{lemma}{Lemma}

\newtheorem{thm}{Theorem}
\newtheorem{prop}{Proposition}
\newtheorem{cor}{Corollary}
\newtheorem{assum}{Assumption}
\newtheorem{defn}{Definition}
\newtheorem{exa}{Example}
\newtheorem{rem}{Remark}

 \allowdisplaybreaks
\def\BibTeX{{\rm B\kern-.05em{\sc i\kern-.025em b}\kern-.08em
    T\kern-.1667em\lower.7ex\hbox{E}\kern-.125emX}}
\makeatletter
\def\fnum@figure{\textcolor{subsectioncolor}{\sf Fig.~\thefigure}}
\def\fnum@table{\textcolor{subsectioncolor}{\sf TABLE~\thetable}}
\def\thm@space@setup{%
  \thm@preskip=3pt
  \thm@postskip=\thm@preskip 
}
 \xpatchcmd{\@thm}{\thm@headpunct{.}}{\thm@headpunct{}}{}{}
\makeatother

\begin{document}
\title{Incentive Mechanisms for Welfare Maximization Under Individual Rationality in Noncooperative Dynamical Systems}
\author{
Yuyue~Yan,  \textcolor[rgb]{0,0,0}{\IEEEmembership{Member,~IEEE}},
Maojiao~Ye, \IEEEmembership{Senior Member,~IEEE},\\
 Hideaki Ishii, \IEEEmembership{Fellow,~IEEE},
and Tomohisa~Hayakawa,  \IEEEmembership{Member,~IEEE}
\thanks{This work was supported in part by JSPS under Grant-in-Aid for
Scientific Research Grant No. 22H01508, by JST
under ASPIRE Grant No. JPMJAP2402, and by the National Natural Science Foundation of China (NSFC) under Grant 62222308, 62173181, 62073171.}
\thanks{Y. Yan and H. Ishii 
are with the Department of Information Physics and Computing, The University of Tokyo, Bunkyo, Tokyo 113-8656, Japan (e-mail: {yan-yuyue@g.ecc.u-tokyo.ac.jp}, hideaki\_ishii@ipc.i.u-tokyo.ac.jp).   }
\thanks{M. Ye is with the School of Automation, Nanjing University of Science and Technology, Nanjing 210094, P.R. China  (e-mail: {ye0003ao@e.ntu.edu.sg}).}
\thanks{T. Hayakawa is with the Department of Systems and Control Engineering, Institute of Science Tokyo, Meguro, Tokyo 152-8552, Japan (e-mail: {hayakawa@sc.e.titech.ac.jp}). }
\vspace{-30pt}}

\maketitle
\begin{abstract}                          
Human societies are  shaped by strategic interactions among self-interested  decision-makers, where conflicts between individual and social interests naturally arise.
This paper investigates the design of incentive mechanism   for multi-agent game systems to reconcile social welfare maximization with individual benefits under economic constraints.
We develop  a   linear-quadratic incentive framework to reshape  Nash equilibria  while reconciling the constraints of Pareto efficiency, individual rationality, and budget balance. 
The optimal target state     is   identified   as the solution of a constrained optimization problem associated with the social welfare  and aggregated payoff functions.
Our theoretical analysis characterizes the conditions  of incentive functions that steer self-interested agents towards the optimal target state as a  Pareto-efficient Nash equilibrium for achieving  the maximum attainable social welfare. 
We further characterize how the choice of social welfare function constrains achievable     individual outcomes, revealing a structural paradox associated with priority weighting under individual rationality. 
Several numerical examples are presented to demonstrate the efficacy of our results.
\end{abstract}

\vspace{-10pt}
\section{Introduction}\label{intro} 
As  future smart  societies are  shaped by strategic interactions among self-interested decision makers,  {it is   important  to  study the coordination issue of       multi-agent game systems}. 
With their self-interested nature,  agents seek   to enhance their own profits by adopting rational strategies. Such pursuit culminates in the establishment of the concept of a Nash equilibrium  in  games \cite{yan2021Stability}. 
For  noncooperative games with dynamic agents,  
 a number of decision dynamics have been developed in the literature  to capture agents' selfish and myopic decision-making processes driven by their payoff structures.
For example, the pseudo-gradient dynamics  \cite{Rosen1965Existence} and  their notable   variants   involving  passive compensator \cite{Li11180003} and differential privacy \cite{cheng2025privacy}    capture  the agents' dynamic  decision behaviors      and  have been    effectively    applied to solve   practical  problems \cite{nekouei2016performance,Gadjov2017APassivity,romano2019dynamic,cortes2015self}.

However, as   Nash equilibria are  selfish solutions of agents in games, 
 it became important to
capture  how  conflicts  between individual and social interests  can be  coordinated  by    incentive-based  regulations \cite{Zhong11370024}. 
To handle the   degradation  in the~social utility (e.g.,  the tragedy of the commons  \cite{hardin1968tragedy})  that can arise  in    noncooperative   systems,  several explicit incentive mechanisms have been imposed externally from a system regulator, who regulates the entire system by changing  agents’ decision-making tendencies.
It is notable that a dynamic pricing mechanism introduced by \cite{Alpcan2009A} achieves the  highest quality-of-service in optical power control  problems. 
Following the idea of aligning    Nash equilibria with the socially optimal state, a large body of research has emerged on    incentive  regulation  \cite{fabiani2023incentives,ratliff2020adaptive,yan2021hierarchical,brown2020can,yan2023pareto}.
For instance, 
the effectiveness of taxes and subsidies was investigated    in \cite{ferguson2021effectiveness} to understand the influence  to the performance of    congestion games.
 Fabiani and Simonetto \cite{fabiani2023incentives} proposed a light-touch policy to incentivize or penalize the service providers   to steer a networked system towards a targeted outcome,
whereas Riehl  \emph{et al.}~\cite{riehl2018incentive}  considered a coordination game  to study   how   binary decision networks can be efficiently driven to a desired equilibrium state by offering payoff incentives or rewards for using a particular strategy.
  \textcolor[rgb]{0,0,0}{While designing the incentive mechanism, the system regulator may have to follow  a sustainable budget  to  reduce the financial pressure on the government \cite{colini2016approximately}.}
In this context, Yan and Hayakawa studied several zero-sum incentive mechanisms to stabilize a possibly unstable Nash equilibrium  \cite{yan2021Stabilization}  and ensure Pareto-improving trajectories \cite{yan2023pareto} under    pseudo-gradient dynamics with unknown    sensitivity parameters.
  \textcolor[rgb]{0,0,0}{An} intra-group and inter-group incentive framework was  presented in  \cite{yan2021hierarchical} to solve a social welfare improvement problem for large-scale  noncooperative dynamical systems.

However, when incentive mechanisms are introduced, economically motivated constraints imposed on the regulator    \textcolor[rgb]{0,0,0}{can fundamentally limit the achievable level of social welfare and therefore require careful consideration.}
{It should be   noticed that}
yielding  a Pareto-efficient Nash equilibrium is one of the typical economic desires  to guarantee  that there is no further Pareto improvement \cite{yan2023pareto}.
Seeking Pareto-efficient Nash equilibria has also  appeared in  engineering applications for optimal bidding decision and efficient resource utilization, such as  \textcolor[rgb]{0,0,0}{those in} relay power allocation \cite{wu2014auction}, edge computing networks \cite{asheralieva2025dynamic}, and wireless sensor networks \cite{zheng2016game}.
On the other hand, in reality, most incentive mechanisms   fail to ensure   that all participants remain free from losses, violating the principle of individual rationality \cite{zhang2025incentivizing} in mechanism design. Such schemes risk dampening participants’ willingness to cooperate with the system regulator. 
Particularly in  open systems with non-coercive incentives,  agents may simply withdraw from competition if their expected gains are  \textcolor[rgb]{0,0,0}{likely to be} unmet  \cite{von2023withdrawal}.
Despite extensive  \textcolor[rgb]{0,0,0}{progresses} in game theory, only a handful of studies have sought incentive mechanisms that  \textcolor[rgb]{0,0,0}{simultaneously} reconcile social welfare maximization, individual rationality, and budget balance. Indeed, theoretical impossibility results \cite{serizawa1996strategy} have shown that these goals cannot be achieved together in certain classes of  \textcolor[rgb]{0,0,0}{Bayesian games.}

{Central to our interest  is an unresolved question:
Can incentive mechanisms align social efficiency with individual benefits without inflicting losses on any participants?
More specifically,  how can we systematically construct incentive mechanisms  that achieve Pareto efficiency and  social maximization  without making any individual worse off?
We must emphasize that in this paper, 
we will fully address these  questions   for general 
concave games, which have  not been studied in the context of mechanism design 
\cite{riehl2018incentive,colini2016approximately,fabiani2023incentives,ferguson2021effectiveness,ratliff2020adaptive,yan2021hierarchical,brown2020can,yan2021Stabilization,yan2023pareto}.}

Specifically, we develop an    {individually} rational incentive mechanism    for  pseudo-gradient-based 
noncooperative dynamical systems where agents act as autonomous decision-makers under  the regulated payoff functions.
Our objective is to establish the theoretical foundations of the designing principle of such mechanisms  and the inherent relationship between the maximum attainable social welfare and the choice of social welfare function. 
In our framework,  the system regulator acts as a financial mediator to   transfer   monetary payoffs  among   agents 
with balanced budget. 
Given the social welfare function and the aggregated welfare function, adaptive controller-based updating laws are proposed for linear-quadratic incentive functions  to change the  decision tendencies of  agents so that both the Nash equilibrium and the states of agents are steered to an optimal target state for social welfare maximization. 
We reveal  that the  target Nash equilibrium is  associated with the solution of  constrained optimization problems \textcolor[rgb]{0,0,0}{due to the constraint of individual rationality}.
We present some conditions to ensure that the Nash equilibrium is shifted towards the optimal target.
Furthermore, sufficient conditions are established to guarantee the convergence of agents’ states to the optimal target   state as a Pareto-efficient Nash equilibrium  under balanced budget.
It is shown  that implementing linear-quadratic incentives is sufficient to  incentive mechanism for the system consisting of more than two agents.

Different from the preliminary version \cite{yan2026cdc},  this paper offers additional  illustrative numerical examples, and detailed proofs.  
Without restricting to  a weighted social welfare function,   this paper focuses on   general social welfare functions and extends    the results of the linear incentive functions in  \cite{yan2026cdc} into linear-quadratic functions to cover the case where the  noncooperative system is not  a strongly monotone game. The existence of  such linear-quadratic incentives is addressed.  

 Furthermore, we  demonstrate the  applicability of our  framework to   networked games of  research and development (R\&D) partnerships and differentiated oligopoly markets with either Cournot or Bertrand competitions. 
Our numerical examples  reveal   an interesting paradox associated with priority weighting in the social welfare function: Prioritizing one agent can eliminate the possibility of benefiting anyone. 
It turns out that the proposed incentive mechanisms can serve as a moderate policy tool to reshape market structures (e.g.,  to influence the commodity prices  or market shares of companies) and enhance collective efficiency without imposing individual losses.

The paper is organized as follows. The incentivized noncooperative system is explained in Section~\ref{section2} with the problem formulation. In Section~\ref{section4}, we  design a framework of incentive mechanisms to achieve social welfare maximization    with convergence analysis. Several numerical examples are  shown in Section~\ref{sec:extension}. 
Finally, we conclude our paper in Section~\ref{section6}.

\emph{Notations}. In this paper, $\mathbb{R}$, $\mathbb{R}_+$, $\mathbb{R}^{n}$, and $\mathbb{R}^{n \times m}$ denote   the sets of real numbers, positive real numbers,  $n\times 1$  real column vectors,  and $n\times m$ real matrices, respectively. 
Moreover,  $(\cdot)^\mathsf T $, ${\rm null}(\cdot) $,  and $\rho(\cdot)$ denote    the transpose,   the null space,   and the spectral radius of a matrix, respectively.
 Furthermore, $[v]_i$ and $[A]_{ij}$ denote the $i$th and $ij$th entry of the vector $v$ and the matrix $A$, respectively.
  For a vector    $v\in\mathbb R^n$, ${\rm diag}[v]$ denotes a diagonal matrix with $ii$th entry being   $[v]_i$.
We write $e_i$ for  $i$-th standard (column) basis, \textcolor[rgb]{0,0,0}{and ${\bf 0}_n$, ${\bf 0}_{n\times n}$, ${\mathds 1}_{n}$,  and ${ I}_{n}$ for an $n\times 1$ zero vector, $n\times n$ zero matrix},     $n$-dimensional ones vector, and the identity matrix, respectively.
  For a set $\mathcal S$, ${\rm int}\, \mathcal S$ and  ${\partial}\mathcal S$ denote its interior and  boundary, respectively. 
Let $|\cdot|$ denote the cardinality of a set. We write  $ \odot$ and  $\oslash$ for Hadamard product and  division. 

We      define     $\Delta_+:=\{\omega\in\mathbb R_+^n:\omega_1+\cdots+\omega_n=1\}$   and further define $\Delta_+^{lr}:=\{\omega\in\Delta_+:|\{\, i\in\mathcal I : \omega_i = \omega_j\}|\le\; n-2,j\in\mathcal I\}$ as  the set of normalized positive weights where  no   weight $w_j$ repeats  more than $n-2$ times.
Let  $[z]_\lambda^+:=z$ if $z>0$ or $\lambda>0$ and $[z]_\lambda^+:=0$ otherwise. 
Finally,   
for a function $f(x)$, let $\partial f(x)$,  $\nabla f(x)$, and  $\nabla^2 f(x)$ denote its  \textcolor[rgb]{0,0,0}{subgradient},  gradient, and  Hessian matrix at $x$ with $\nabla_{\!i} f(x):={ \partial  f(x)}/{\partial x_i}$\textcolor[rgb]{0,0,0}{, respectively}.

 \vspace{-6pt}
\section{Noncooperative Systems and Problem Formulation}\label{section2}\vspace{-3pt}
\subsection{System Description}
Consider an  $n$-agent noncooperative system supervised by  a  system regulator, where the set of  {agent indices} is denoted by $\mathcal{I}:=\{1,\dots,$ $n\}$.
Each agent $i\in\mathcal{I}$   controls its state (strategy) $x_{i} \in \mathbb{R}$ to increase  its   payoff function denoted by $f_i:\mathbb R^n\to\mathbb R: x \mapsto f_i(x) $, 
where  $x:=[x_1,\ldots,x_n]^{\mathsf T  }=(x_i, x_{-i})\in\mathbb R^n$ denotes  the state profile of overall    agents, and 
$x_{-i}$ is the state profile except agent $i$. The system regulator     cannot directly control decisions of agents' strategies, but can~design  a monetary \emph{incentive}  function  $y_i:\mathbb R^n\to\mathbb R$  as either  taxes or subsidies for agent $i$ based on the knowledge of the state profile   $x$ of all agents to enhance the  collective~outcome  represented by a  predefined social welfare function $\mathcal W:\mathbb R^n\to\mathbb R$. 
We introduce two concepts of social welfare candidates   as follows: 

\emph{(a) Utilitarian social welfare}: The utilitarian (also known as Benthamite or aggregated) social welfare function \cite{sen1977weights,etesami2024maximizing} is defined by
\begin{equation}
\mathcal W_{\rm u}(x):= \sum_{i\in\mathcal I}f_i(x).
\end{equation}

\emph{(b) Weighted social welfare}: The weighted social welfare function \cite{etesami2024maximizing}   is defined by 
  \begin{equation}\label{weijt_f}
\mathcal W_{\rm w}(x):= \sum_{i\in\mathcal I}\xi_if_i(x),
\end{equation}
where $\xi=[\xi_1,\ldots,\xi_n]^\mathsf T\in \Delta_+$ are  normalized positive welfare  weights   representing   policy priority  of agents $i\in\mathcal I$  predefined  by the system regulator.  
 
We consider the situation where  no external financial  capital can be injected or extracted and thus the system regulator~follows the so-called strong balanced budget  \cite{colini2016approximately,murhekar2023incentives}     given by
\begin{equation}\label{bdgt}
y_1(x)+y_2(x)+\cdots+y_n(x)=0,\quad x\in\mathbb R^n.
\end{equation}
This relation    expresses the role of the   system regulator being purely  a mediator, implementing monetary transfers (taxes or subsidies) among the agents.
After accounting for the transfers, the \emph{effective} (modified) payoffs of  agents are defined as the net payoffs given by
\begin{equation}\label{netw}
V_i(x):= f_i(x)+y_i(x), \quad i\in\mathcal I,
\end{equation}
where the incentive terms are positive for subsidies, negative for taxes, and zero when no transfer occurs. 
Henceforth, the  un-incentivized and incentivized noncooperative  systems
are  denoted by  $\mathcal G(f)$ and $\mathcal G(V)$, respectively. 
The definitions of Nash equilibrium and Pareto efficiency are introduced below.

 \begin{defn} \label{def1}  For the incentivized  noncooperative game $\mathcal G(V)$, the state profile $\tilde x^{*}\in \mathbb R^n$ is called a Nash equilibrium if
\begin{equation}\label{def:NE}
V_i(\tilde x_i^*,\tilde x_{-i}^*)\geq V_i( x_i,\tilde x_{-i}^*),\quad  x_i\in\mathbb{R},\quad i\in\mathcal{I}.
\end{equation}
\end{defn}

\begin{defn}  
The state profile  $\tilde x^*\in\mathbb R^n$ is said to be Pareto efficient in the  incentivized   noncooperative game   if
no  strategy $x\in\mathbb R^n$ exists such that $V_i(x)\geq  V_i(\tilde x^*)$ for all $i\in\mathcal I$  with at least one strict inequality   for some $i\in\mathcal I$.
\end{defn}

Although a non-Pareto-efficient Nash equilibrium may admit Pareto improvements, no agent has any intention to deviate unilaterally. Therefore, the system regulator must design incentive mechanisms that shift the Nash equilibrium to a Pareto-efficient equilibrium of $\mathcal G(V)$ with higher social welfare $\mathcal W(x)$ than that at the original Nash equilibrium (which is denoted by  $x^*\in\mathbb R^n$ henceforth) of $\mathcal G(f)$. 
Note that Pareto-efficient Nash equilibria are desirable in many practical  systems, including edge computing  \cite{asheralieva2025dynamic}, relay power allocation \cite{wu2014auction},  and wireless sensor networks \cite{zheng2016game}, where efficient resource allocation and optimal bidding decisions are required.

{The following assumptions are used in the paper to restrict the noncooperative system $\mathcal G(f)$  to concave games  \cite{Rosen1965Existence}.}
  \begin{assum}\label{as_Ji}
\emph{
Each  payoff function  $f_i(x)$ is twice continuously differentiable   and strictly concave in $x_i$ for all  $i\in\mathcal{I}$.}
 \end{assum}
 
   \begin{assum}\label{as_Ji2}
\emph{
The gradients of   payoff functions  $f_i(x)$, $i\in\mathcal{I}$, are  globally Lipschitz.}
 \end{assum}
 
 \begin{assum}\label{as_social}
\emph{The  social welfare function $\mathcal W(x)$  is strictly  concave with respect to $x$.}
\end{assum}

 Under Assumption~\ref{as_Ji}, when the incentive functions are also  continuously differentiable, it holds that  a state $\tilde x^*\in\mathbb R^n$ is a Nash equilibrium of $\mathcal G(V)$  only if 
\begin{equation}\label{gradient}
\frac{\partial  V_i(\tilde x^*)}{\partial x_i}=\frac{\partial  f_i(\tilde x^*)}{\partial x_i}+\frac{\partial  y_i(\tilde x^*)}{\partial x_i}=0, \quad i\in\mathcal{I}.
\end{equation}
The Nash equilibrium is  contingent on the incentive functions, but is characterized independently of the underlying dynamics of agents.

\vspace{-10pt}
\subsection{Dynamic Agents With Reactive Decision-Making}
Each agent is assumed to know only its own (regulated)~payoff function, with no information about how others’ payoffs are determined.
This  fact makes each agent~$i$ an autonomous and myopic decision maker who continually updates its individual
state $x_i(t)$ by acting  in response to others’ strategies $x_{-i}(t)$ at time $t\geq0$.
Although no agent has the knowledge of the Nash equilibria, their self-improving and short-sighted behavior may still lead the group toward one. 
Assuming that  the state profile    $x(t)$ is available for all  agents,  we  formulate the behavioral dynamics of the agents     as the pseudo-gradient dynamics \cite{Rosen1965Existence} 
\begin{equation} \label{pdg}
\dot{x}_i(t)= \beta_i\nabla\!_i V_i(x_i(t),x_{-i}(t)),\ \  i\in\mathcal{I}, 
\end{equation}
for an initial state $ x(0)=x_0\in\mathbb R^n$, where $\beta_i\in\mathbb R_+$, $i\in\mathcal I$,  are  agent-dependent positive  parameters   representing sensitivity   to the increasing/decreasing payoff per unit strategy change \cite{hurst2017cooperation,yan2024Incorporation}.  
Clearly, from \eqref{gradient},  any Nash equilibrium $\tilde x^*$ of $\mathcal G(V)$  is an   equilibrium point  of the pseudo-gradient  dynamics \eqref{pdg}. For our analysis,  we write  the   dynamics \eqref{pdg} in vector form    as 
\begin{equation}\label{dynamics2}
\dot{x}(t)={\rm diag}[\beta] (F(x(t))+  \Phi (x(t))),
\end{equation} 
where   $F(x):=[\nabla_{\!1} f_1(x),\ldots,\nabla_{\!n} f_n(x)]^{\mathsf T}\in\mathbb R^n$ and $\Phi(x):=[\nabla_{\!1} y_1(x),\ldots,\nabla_{\!n} y_n(x)]^{\mathsf T}\in\mathbb R^n$  are the
pseudo-gradient functions of agents' payoff   and  incentive functions, respectively, and $\beta:=[\beta_1,\ldots,\beta_n]^\mathsf T\in\mathbb R_+^n$.
\vspace{-8pt}
\subsection{Motivations and Problem Statement} 
Before we  present the main problem of this paper, we discuss  some motivations of this work. 

For non-coercive incentive mechanisms,  
agents (companies) may exit the market when their payoffs of equilibrium outcome  decrease after the incentive mechanism is executed.
Therefore, it is crucial
to       design an incentive mechanism  \textcolor[rgb]{0,0,0}{ensuring that} each agent's payoff at the equilibrium outcome is improved by the mechanism.
An incentive mechanism achieving \textcolor[rgb]{0,0,0}{monotonically} increasing trajectories of payoffs  is developed in  \cite{yan2023pareto}.
 However,  we would like to point out that neither this approach nor   those in \cite{Alpcan2009A,shakarami2023dynamic,maheshwari2022inducing,murhekar2023incentives,yan2021hierarchical,yan2025social,chattopadhyay2019nash,yan2025Information}  may   strictly enhance  the  equilibrium outcome by employing  the incentive mechanism. 
To retain individual benefits for all participants together with collective outcomes, we introduce below  the notion of an \emph{individually rational Pareto-efficient}  incentive mechanism   with respect to a baseline equilibrium  $x_{\rm bs}\in\mathbb R^n$. For example, this baseline  can be taken as the Nash equilibrium $x^*$ of   $\mathcal G(f)$ without any regulation ($y_i(x)\equiv 0$, $i\in\mathcal I$).

\begin{defn}\label{Def:IRPE}
Compared to a  baseline equilibrium $x_{\rm bs}\in\mathbb R^n$, an incentive mechanism \eqref{netw}  is said to be {individually rational  Pareto efficient} (IRPE)\footnote{Notice  that when the  baseline equilibrium $x_{\rm bs}$ is taken   as the initial state $x_0$, the IRPE mechanism with $y_i(x_0)=0$, $i\in\mathcal I$,   is  understood as      a steady-state version of Pareto-improving incentive mechanisms defined in \cite{yan2023pareto}.  }  if there is a  Pareto-efficient Nash equilibrium $\tilde x^*$ such that 
  \begin{align}\label{def:C3}
V_i(\tilde x^*)\geq   f_i( x_{\rm bs}),\quad i\in\mathcal I.
  \end{align}
Furthermore, the incentive mechanism  is  called {optimal IRPE} if  the Pareto-efficient Nash equilibrium  $\tilde x^*$ further maximizes the social welfare  $\mathcal W(x)$ subject to  individual rationality   \eqref{def:C3}.
\end{defn}

In Definition~\ref{Def:IRPE},  the condition \eqref{def:C3} is the individual rationality condition\footnote{Note that the concept of IRPE incentive mechanism is  distinct from related notions in classical mechanism design \cite{matsushima2019efficient}, which is typically developed for incomplete-information games with     Bayesian Nash equilibrium.}; it    requires that   the modified equilibrium outcome of $\tilde x^*$  guarantees every agent a performance  level   at least as high as that  at the baseline equilibrium $x_{\rm bs}$. In other words, when the baseline equilibrium  is set to $x_{\rm bs}=x^*$ for the   system $\mathcal{G}( f)$,
  no agent should incur a loss after the  regulation (even though some agents may be required to pay taxes).
It should be also noted  that the constraints of Pareto efficiency, individual rationality, and balanced budget of incentive mechanisms may fundamentally limit the 
 achievable level of social welfare relative to the socially optimal state $\hat x^{\rm opt}$ defined  by
\begin{equation}\label{xopt}
\hat x^{\rm opt}:= \arg\max\nolimits_{x\in\mathbb R^n}\mathcal W(x),
\end{equation}
which was   the target state in most of the literature \cite{Alpcan2009A,yan2025social,yan2025Information}. 
Now, we present the problem of this paper as follows.

\subsubsection*{Problem}
Consider the noncooperative system $\mathcal{G}(V)$ with the pseudo-gradient dynamics \eqref{pdg}. Given the  social welfare function $\mathcal W(x)$, 
the main objective of this paper is two-fold: i) Identify the optimal target state  $\hat{\mathbf x}\in\mathbb R^n$  for the optimal IRPE  incentive mechanism. ii) Design the optimal IRPE  incentive functions $y_i(x)$, $i\in\mathcal I$, satisfying the  balanced budget     \eqref{bdgt}
to ensure that the agents' state  $x(t)$ converges to an optimal  target state   $\hat{\mathbf x}$ as a Pareto-efficient Nash equilibrium in $\mathcal{G}(V)$.

\section{Achieving Social Welfare Maximization Under   Optimal IRPE Incentive Mechanisms}\label{SecIIIA} \label{section4} 
In this section, we characterize an optimal IRPE  incentive mechanism  for the noncooperative system  $\mathcal G(f)$ under strong balanced budget \eqref{bdgt} with   linear-quadratic functions.
{Specifically, we consider the incentive functions of the form}
\begin{equation}\label{incen_infla}
y_i(x)=  \tfrac{1}{2}\|x-x_{\rm r}\|_{K_i}^2+u_i(x-x_{\rm r})+\tau_i(x_{\rm r}), \quad i\in\mathcal I,
\end{equation}
where $x_{\rm r}\in\mathbb R^n$ is a time-varying reference signal generated by the system regulator; this converges to an optimal target state $\hat{\mathbf x}$ as will be analyzed  later. 
Moreover, $K_i\in\mathcal K_i:=\{{\rm diag}[k_i^1,\ldots,$ $k_i^n]\in\mathbb R^{n\times n}: k_i^i\leq 0\}$  is a constant Hessian matrix, 
$u_i=[u_i^1,\ldots,u_i^n]\in\mathbb R^n$ denotes  a time-varying vector, and $\tau_i(x_{\rm r})$ denotes a scalar function. 
{For the strong balanced budget \eqref{bdgt}, we assume that} 
\begin{equation}\label{balanced_KU}
\sum_{i\in\mathcal I} { K}_i={\bf 0}_{n\times n},\ \ \sum_{i\in\mathcal I} { u}_i(t)={\bf 0}_{n},\ \  \sum_{i\in\mathcal I} \tau_i(x_{\rm r}(t))=0,
\end{equation}
for  $t\geq 0$. The next assumption is used in the rest of the paper. 

 \begin{assum}\label{as_u_i}
 \emph{The updating rules of the reference signal  $x_{\rm r}(t)$ and the   vector $ u_i(t)$ do not depend on $x_i(t)$, i.e., ${ \partial  x_{\rm r}}/{\partial x_i} ={\bf 0}_n$ and ${ \partial  u_i}/{\partial x_i} ={\bf 0}_n$ for      $i\in\mathcal{I}$.}
  \end{assum}
  
Under Assumption~\ref{as_u_i},  
 the  pseudo-gradient dynamics \eqref{pdg} of the incentivized noncooperative system $\mathcal G(V)$ are given by
\begin{align}\label{dyan_closed}
\dot x (t)={\rm diag}[\beta]\left[ F(x(t))+u(t)+K(x(t)-x_{\rm r}(t))\right],
\end{align}
where $u:=[u_1^1,\ldots, u_n^n]^{\mathsf T}\in\mathbb R^n$ and $K:={\rm diag}[k_1^1,\ldots,k_n^n]\leq 0$ are characterized from the pseudo-gradient of the linear\footnote{{Note that $u$ contains only $u_i^i$ from $u_i$, $i\in\mathcal I$. The rest    in $u_i$, $i\in\mathcal I$, are used to guarantee  Pareto efficiency and balanced budget as discussed in Sections~\ref{swcaaaa} and \ref{nonequal_sin}}.} and quadratic terms of the incentive functions \eqref{incen_infla}, respectively.
Furthermore,   $y_i(x)$ is continuously differentiable  and concave in $x_i$ for   $i\in\mathcal{I}$; this implies together  with Assumption~\ref{as_Ji} that the Nash equilibria of $\mathcal G(V)$  are identical to the equilibria of the pseudo-gradient dynamics \eqref{pdg}. 
As a result, the  linear terms of \eqref{incen_infla} contribute to shifting the Nash equilibria of $\mathcal G(V)$ and the quadratic terms contribute to  changing the Jacobian matrix of the vector field and thus stabilizing the closed-loop dynamics. In particular, for yielding a Nash equilibrium at $\hat{\mathbf x}$, the control input $u(t)$ should satisfy
\begin{equation} \label{ui_cond}
u(t)\to u^*:=-F(\hat{\mathbf x})  \ {\rm as}\  t\to\infty. 
\end{equation}
  We notice that the incentive functions  \eqref{incen_infla} are  different from those in   \cite{yan2021Stabilization,yan2025social} due to the additional enforced scalar   $\tau_i(x_{\rm r})$, which never changes the pseudo-gradient of the noncooperative system but contributes   to ensuring   individual rationality as we will discuss later. 
 
 Before we present the detailed updating laws of $x_{\rm r}(t)$, $u_i(t)$, $\tau_i(x_{\rm r})$, $i\in\mathcal I$, we first establish  the  theoretical foundations of the maximum attainable social welfare of IRPE incentives.

\subsection{Characterization of Optimal Target States}\label{swqoo}
In this part, we characterize the optimal target state  $\hat{\mathbf x}$.  
 To this end, we define a utilitarian benefiting region  representing the set of states  whose utilitarian welfare $ {\mathcal W_{\rm u}}(x)$  exceeds that of the baseline equilibrium $x_{\rm bs}$ by 
 \begin{equation}\label{XWusa}
 \mathcal X_{\mathcal W_{\rm u}}:=\{x\in\mathbb R^n:  {\mathcal W_{\rm u}}(x)- {\mathcal W_{\rm u}}(x_{\rm bs})\geq 0\}.
\end{equation}   

 \begin{assum}\label{as_social0}
\emph{The utilitarian social welfare function $\mathcal W_{\rm u}(x)$  is strictly  concave.}
\end{assum}

Assumption~\ref{as_social0} indicates that   the utilitarian benefiting region  $ \mathcal X_{\mathcal W_{\rm u}}$ is non-empty, bounded, and convex. 
The following result shows that the individual rationality condition  \eqref{def:C3}  restricts the admissible target state $\hat{\bf x}$ to this region, indicating that    optimal level  of social performance    $\mathcal W(\hat x^{\rm opt})$  may not be attained when  $\hat x^{\rm opt}\not\in \mathcal X_{\mathcal W_{\rm u}}$.

\begin{lemma}\label{Lema1}The target state   $\hat{\bf x}$  of the   IRPE  incentive mechanism with strong balanced budget \eqref{bdgt} must satisfy  $\hat{\bf x}\in \mathcal X_{\mathcal W_{\rm u}}$.   
\end{lemma} 

\noindent\textbf{Proof} The individual rationality condition \eqref{def:C3}  implies $V_i( \hat{\bf x})\geq   f_i( x_{\rm bs})$ and thus it follows from summing the effective payoffs \eqref{netw} under  \eqref{bdgt} at the  target state $\hat{\bf x}$ that
  \begin{align}\nonumber 
 \sum_{i\in\mathcal I} V_i( \hat{\bf x})=& \sum_{i\in\mathcal I} f_i( \hat{\bf x})+  \sum_{i\in\mathcal I} y_i( \hat{\bf x}) \allowdisplaybreaks \\
=& \sum_{i\in\mathcal I} f_i( \hat{\bf x}) \geq \sum_{i\in\mathcal I} f_i( x_{\rm bs}).
  \end{align}
 Hence, $\hat{\bf x}$ is an equilibrium under IRPE  incentives only if  $\hat{\bf x}\in \mathcal X_{\mathcal W_{\rm u}}$. The proof is complete. 
  $\hfill\square$

\subsubsection{Necessary and Sufficient   Condition of Individual Rationality} The next result  reveals   that the implementation of~individual rationality   \eqref{def:C3}  with the strong balanced budget \eqref{bdgt}   depends solely on steady-state payoff transfers,  which are decoupled to the implementation of Pareto-efficient Nash equilibrium.  

\begin{prop}\label{lemma2aq} The incentive mechanism \eqref{netw} satisfying a strong balanced budget \eqref{bdgt} is  individually rational  at the  target state  $\hat{\bf x}\in \mathcal X_{\mathcal W_{\rm u}}$, i.e., $V_i( \hat{\bf x})\geq   f_i( x_{\rm bs})$, $i\in\mathcal I$, if and only if  
 \begin{equation}\label{offest_eq}
 y_i(\hat{\bf x})=\eta_i[\mathcal W_{\rm u}(\hat{\bf x})-\mathcal W_{\rm u}(x_{\rm bs})]+f_i(x_{\rm bs})-f_i(\hat{\bf x}),\ i\in\mathcal I,
\end{equation}  
for some    $\eta_i\geq0$, $i\in\mathcal I$, satisfying  $\sum_{i\in\mathcal I}\eta_i=1$.
\end{prop} 

\noindent\textbf{Proof} For sufficiency,  it follows from \eqref{offest_eq} that 
  \begin{align}\nonumber
V_i(\hat{\mathbf x})&=f_i(\hat{\mathbf x})+y_i(\hat{\mathbf x})\\ \label{Ssqd}
&=\eta_i[\mathcal W_{\rm u}(\hat{\bf x})-\mathcal W_{\rm u}(x_{\rm bs})]+f_i(x_{\rm bs}),\quad i\in\mathcal I,
  \end{align}
  which implies $V_i( \hat{\bf x})\geq   f_i( x_{\rm bs})$ for all agents $i\in\mathcal I$, since $\hat{\bf x}\in \mathcal X_{\mathcal W_{\rm u}}$ implies  $\mathcal W_{\rm u}(\hat{\bf x})-\mathcal W_{\rm u}(x_{\rm bs})\geq 0$.
  
   For necessity, note from Lemma~\ref{Lema1}   that there are only two cases, that is,  $\hat{\mathbf{x}}\in\partial\mathcal X_{\mathcal W_{\rm u}}$ and $\hat{\mathbf{x}}\in{\rm int}\,\mathcal X_{\mathcal W_{\rm u}}$. First, consider the case where $\hat{\mathbf{x}}\in\partial\mathcal X_{\mathcal W_{\rm u}}$ holds, i.e.,  $\mathcal W_{\rm u}(\hat{\bf x})-\mathcal W_{\rm u}(x_{\rm bs})=0$. In this case, \emph{ad absurdum}, assume that  \eqref{offest_eq} does not hold.   It then  follows that  
       \begin{equation}  
 \sum_{i\in\mathcal I} y_i(\hat{\bf x})\!\not=\!\sum_{i\in\mathcal I} f_i(x_{\rm bs})\!-\!\!\sum_{i\in\mathcal I} f_i(\hat{\bf x})=0,\end{equation} 
 which contradicts with the strong balanced  budget  \eqref{bdgt}. Hence,  the expression  \eqref{offest_eq} is a necessary condition. 
  Next, consider the case where  $\hat{\mathbf{x}}\in{\rm int}\,\mathcal X_{\mathcal W_{\rm u}}$, i.e.,  $\mathcal W_{\rm u}(\hat{\bf x})-\mathcal W_{\rm u}(x_{\rm bs})>0$. In this case, it holds   from \eqref{offest_eq}  that $\sum_{i\in\mathcal I}\eta_i=1$     since 
   \begin{align}  
0&= \sum_{i\in\mathcal I} y_i(\hat{\bf x})=(\sum_{i\in\mathcal I}\eta_i-1) [\mathcal W_{\rm u}(\hat{\bf x})-\mathcal W_{\rm u}(x_{\rm bs})].
\end{align} 
Then, the proof is immediate since $\eta_i\geq0$, $i\in\mathcal I$, are necessary to guarantee $V_i( \hat{\bf x})\geq   f_i( x_{\rm bs})$, $i\in\mathcal I$, in \eqref{Ssqd}.    $\hfill\square$

\begin{rem}\label{remal:1}
\emph{Note that the utilitarian surplus  ${\mathcal W_{\rm u}}(\hat{\bf x})- {\mathcal W_{\rm u}}(x_{\rm bs})$ indicates the possibility of strict payoff improvement of agents.  Furthermore,  the condition  \eqref{offest_eq} is     a general steady-state payoff redistribution policy  to allocate the utilitarian surplus  $\mathcal W_{\rm u}(\hat{\bf x})-\mathcal W_{\rm u}(x_{\rm bs})$  towards each agent  $i$ with the allocation  ratio  being $\eta_i$.  Under the mechanism, all agents are strictly better off than at the baseline equilibrium $x_{\rm bs}$ with positive  allocation  ratios  if  $\hat{\mathbf{x}}\in{\rm int}\,\mathcal X_{\mathcal W_{\rm u}}$; otherwise, agents retain their baseline payoffs $V_i(\hat{\mathbf{x}})=f_i(x_{\rm bs})$,      $i\in\mathcal I$ (for a more detailed discussion of the optimal target state $\hat{\bf x}$, see  Example~\ref{exa1} in Section~\ref{swcaaaa}). 
}
\end{rem} 

\subsubsection{Sufficient and Necessary Conditions of Pareto Efficiency}
Proposition~\ref{lemma2aq} above has shown   that the identification of~optimal target state $\hat{\bf x}$ for IRPE incentive mechanisms reduces to checking the   feasibility of    implementing a Pareto-efficient Nash equilibrium at $\hat{\bf x}\in\mathcal X_{\mathcal W_{\rm u}}$. Thus,  we next characterize~the sufficient conditions and necessary conditions of Pareto efficiency. 

\begin{lemma} \label{lemma:sp1}
For a given target state $\hat{\mathbf x}\in\mathbb R^n$, if  
\begin{equation} \label{PENE1}
\hat{\mathbf x}
 \in\arg\max_{x\in\mathbb R^n}\sum\nolimits_{i\in\mathcal I}  w_i V_i(x)
\end{equation} 
for some weight $w=[w_1,\ldots,w_n]^{\mathsf T}\in\Delta_+$, 
then this target state  $\hat{\mathbf x}$  is Pareto efficient   in $\mathcal G(V)$.
\end{lemma}

\noindent\textbf{Proof}
Since any Pareto improvement from  $V_1(\hat{\mathbf x}),\ldots,V_n(\hat{\mathbf x})$ would increase $\sum_{i\in\mathcal I} w_i  V_i(x)$   and contradicts with the condition \eqref{PENE1}, 
the proof is complete. 
 $\hfill\square$

\begin{figure}
\centering
\includegraphics[width=0.65\linewidth]{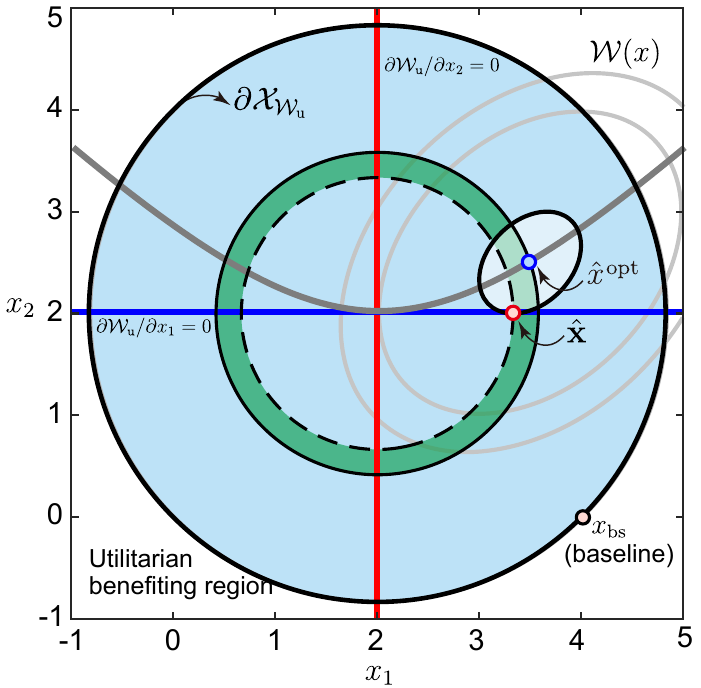}
\caption{{Phase portrait of a two-agent noncooperative system where the socially optimal state $\hat x^{\rm opt}$ is an inadmissible target   since $\hat x^{\rm opt}$ does  not appear on the zero-gradient curves  $\partial {\mathcal W_{\rm u}}(x)/\partial x_i=0$, $i=1,2$.} Note that the grey ellipses, parabola, and red node represent the  level sets of  social  welfare $\mathcal W(x)$, the Pareto front of $\mathcal G(f)$, and the optimal target state $\hat{\mathbf x}$ in $\mathcal G(V)$, respectively. The detailed setting is discussed   in Example~\ref{exa1} in Section~\ref{swqoo}.}
\label{siji1}
\end{figure}

\begin{rem}
\emph{For the utilitarian social welfare function $\mathcal W(x)=\mathcal W_{\rm u}(x)$,  Lemma~\ref{lemma:sp1} implies that  the socially optimal state $\hat x^{\rm opt}$   is Pareto efficient for any budget-balanced  incentive functions  since the condition \eqref{PENE1} holds for $w={\mathds 1}_n/n$. In this case, the  construction of the  optimal  IRPE incentive mechanism~reduces  to a simple Nash equilibrium stabilization problem at $\hat x^{\rm opt}$~with the balanced budget as shown  in \cite{yan2021Stabilization}.  However,  for a non-utilitarian type of social welfare, all directions of the gradient vectors $u_i$, $i\in\mathcal I$, may need to be  controlled simultaneously.  
}
\end{rem}

The next  lemma provides a  necessary condition for implementing IRPE incentives in two-agent noncooperative systems. It indicates  that    an optimal level  of social performance    $\mathcal W(\hat x^{\rm opt})$  may not be attained even when  $\hat x^{\rm opt}\in \mathcal X_{\mathcal W_{\rm u}}$ (see Fig.~\ref{siji1} for an  example of the infeasible target $\hat x^{\rm opt}$  shown as  the blue point).

\begin{lemma} \label{lemma:sp2}
Consider a two-agent noncooperative system $\mathcal G(f)$. Then,   a given  target state $\hat{\mathbf x}\in\mathbb R^n$ of  the incentive mechanism \eqref{netw} satisfying  a strong balanced budget \eqref{bdgt} is  a Pareto-efficient  Nash equilibrium in $\mathcal G(V)$  only if 
 \begin{align}  \label{anlqq}
\hat{\bf x}\in\{x\in\mathbb R^n:\nabla\!_1 {\mathcal W_{\rm u}}(x) \nabla\!_2 {\mathcal W_{\rm u}}(x)=0\},
\end{align}
i.e.,  only if $\hat{\mathbf x}$ appears on   the zero-gradient curve  $\nabla\!_1 {\mathcal W_{\rm u}}(x)=0$ or $\nabla\!_2 {\mathcal W_{\rm u}}(x)=0$.
\end{lemma}

\noindent\textbf{Proof}
{Define the contract curve \cite{varian1992microeconomic}  as
 \begin{equation} 
\mathcal C_{\rm con}:=\!\left\{\!x\in\mathbb R^2\!: \!\frac{\partial V_1(x)}{\partial x_2}\frac{\partial V_2(x)}{\partial x_1}=\frac{\partial V_1(x)}{\partial x_1}\frac{\partial V_2(x)}{\partial x_2} \!\right\}.
\end{equation} 
At a  Pareto-efficient  state $\hat{\mathbf  x}$, due to  continuously differentiability of $V_1(x)$ and $V_2(x)$,        the level sets of $V_i(x)=V_i(\hat{\bf  x})$, $i=1,2$, should be tangent to each other. This  indicates  that   $\nabla V_1(\hat{\bf x})=\eta \nabla V_2(\hat{\mathbf x})$ for  some    scalar $\eta\not=0$. Therefore, the target state  $\hat{\bf x}$ is Pareto efficient  in $\mathcal G(V)$ only if $\hat{\bf x}\in\mathcal C_{\rm con}$. 

Now, since   $\hat{\bf  x}$  is 
a Nash equilibrium, it holds that  $\nabla\!_i y_i(\hat{\bf  x})=-\nabla\!_i f_i( \hat{\bf  x})$ for $i=1,2$. 
Thus, it holds from the balanced budget  \eqref{bdgt} that      
$
\nabla\!_i y_i(   \hat{\bf  x})=-\nabla\!_i y_j(   \hat{\bf  x})=-\nabla\!_i f_i(  \hat{\bf  x})$, $i=1,2$, $j\not=i$.
Then, the necessary condition $ \hat{\bf  x}\in\mathcal C_{\rm con}$ indicates 
 \begin{align} \nonumber
0&=\frac{\partial V_1( \hat{\bf  x} )}{\partial x_2}\frac{\partial V_2(  \hat{\bf  x} )}{\partial x_1} \allowdisplaybreaks \\ \nonumber
&=(\nabla\!_2  f_1(  \hat{\bf  x})+\nabla\!_2  y_1( x^*))(\nabla\!_1  f_2(  \hat{\bf  x} )+\nabla\!_1  y_2(  \hat{\bf  x})) \allowdisplaybreaks \\ \nonumber
&=(\nabla\!_2  f_1(  \hat{\bf  x})+\nabla\!_2 f_2( x^*))(\nabla\!_1  f_2( \hat{\bf  x} )+\nabla\!_1  f_1( \hat{\bf  x})) \allowdisplaybreaks \\ 
&=\nabla_1\,   {\mathcal W_{\rm u}}( \hat{\bf  x})\nabla_2 \,  {\mathcal W_{\rm u}}(  \hat{\bf  x}).
\end{align}
This further  implies    $\nabla_1\,   {\mathcal W_{\rm u}}(  \hat{\bf  x})=0$ or  $\nabla_2\,   {\mathcal W_{\rm u}}( \hat{\bf  x})=0$ at the target state $\hat{\bf  x}$. 
The proof is complete. 
  $\hfill\square$

\subsubsection{Optimal Target States}\label{swcaaaa}
We are now ready to state our first main result on   the optimal target state $\hat{\bf x}$, which maximizes the social welfare function $\mathcal W(x)$ under the constraints of Nash, Pareto efficiency, individual rationality, and balanced budget. 

\begin{thm}\label{thm:main1}
Consider the $n$-agent noncooperative system $\mathcal G(f)$ under  Assumptions~\ref{as_Ji}--\ref{as_social0}.  
Then,  given a social welfare function $\mathcal W(x)$,   the maximum attainable social welfare $\mathcal W(\hat{\bf x})$ of IRPE incentive mechanisms satisfies the following properties:

(a) If $n\geq 3$, then  the optimal target state is given by  \begin{equation} \label{hatx}
\hat{\bf x}=\arg\max_{x\in\mathbb R^n} \mathcal W(x)\ \ 
{\rm s.t.}\,   {\mathcal W_{\rm u}}(x)- {\mathcal W_{\rm u}}(x_{\rm bs})\geq0.
\end{equation}

(b) If $n= 2$, then the optimal target state is given by 
 \begin{align} \nonumber 
\hat{\bf x}=\arg\max_{x\in\mathbb R^n} \mathcal W(x)\ \ 
{\rm s.t.}\, &  {\mathcal W_{\rm u}}(x)- {\mathcal W_{\rm u}}(x_{\rm bs})\geq0,\\ \label{hatx2}
 		&\nabla\!_1\,  {\mathcal W_{\rm u}}(x) \nabla\!_2\,  {\mathcal W_{\rm u}}(x)=0.
\end{align}
\end{thm}
 
 \noindent\textbf{Proof} It follows from Proposition~\ref{lemma2aq} that the individual rationality at $\hat{\mathbf x}$  is  achieved by \eqref{offest_eq} since   $\hat{\bf x}\in \mathcal X_{\mathcal W_{\rm u}}$ holds under \eqref{hatx} and \eqref{hatx2}. 
First, to show the existence of incentive functions for $n\geq3$, let $w=[w_1,\ldots,w_n]^\mathsf T\in\Delta_+^{lr}$ be a positive weighting vector   where no  welfare weight   $w_i$   appears $n-1$
times. The Hessian matrix of  $L(x):=\sum\nolimits_{i\in\mathcal I}  w_i [y_i(x) + f_i(x)]$  in the right-hand side of \eqref{PENE1} is denoted as $\mathbf H\in\mathbb R^{n\times n}$.  Define     
\begin{align}\label{def:taui}
\delta_i:=\sum\nolimits_{s\in\mathcal I} \xi_s \delta_{is}, \quad i\in\mathcal I,
\end{align} which are   finite under Assumption~\ref{as_Ji2} with  $\delta_{is}$ defined by 
\begin{align}\label{def:tauis}
\delta_{is}:=\! -\sup_{x\in\mathbb R^n}\!\!\Big[\nabla_{ii}^2 f_s( x)+\sum_{j\not=i} \big|\nabla_{ij}^2 f_s( x)\big|\Big],\ i,s\in\mathcal I.
\end{align} 
Now, note that under the condition
\begin{align} \label{inequa_kk0}[   k_1^{i} ,  \ldots ,   k_n^i]w\leq   \delta_i ,
\end{align} 
  the Hessian matrix $\bf H$ of the  function $L(x)$ satisfies 
\begin{align}\nonumber 
\!\!\!\![{\bf H}]_{ii}\!&
=\sum_{s\in\mathcal I} w_s[\nabla_{ii}^2 f_s( x)\!+\!\nabla_{ii}^2 y_s(  x)]\leq\!\sum_{s\in\mathcal I} w_s[\nabla_{ii}^2 f_s( x)]\!+\!\delta_i
\\
\nonumber
&=\sum_{s\in\mathcal I} w_s\Big\{\nabla_{ii}^2 f_s( x)\! -\!\!\sup_{x\in\mathbb R^n}\!\!\Big[\nabla_{ii}^2 f_s( x)\!+\!\sum_{j\not=i} \big|\nabla_{ij}^2 f_s( x)\big|\Big]\!\Big\}
\\  
&\leq-\!\sum_{j\not=i}\sum_{s\in\mathcal I}\! w_s\Big|\nabla_{ij}^2 f_s( x)\Big|=\!-\!\sum_{j\not=i}\Big|[{\bf H}]_{ij}\Big|,\  i\in\mathcal I.
\end{align}
 Thus, the matrix  $\bf H$ is strictly diagonally dominant and hence negative definite, indicating  that the function $L(x)$ is concave. 
 
 Recall $k_i^i\leq0$, $i\in\mathcal I$, and the constraint of balanced budget \eqref{balanced_KU}. Thus,     the condition \eqref{inequa_kk0} can be further written as 
 \begin{subequations}\label{inequa_kk}\begin{align}
[w,e_i]^{\mathsf T}&[   k_1^{i} ,  \ldots ,   k_n^i
]^{\mathsf T}\leq[  \delta_i ,0]^{\mathsf T},
\\
 \label{inequa_kk2}
{\rm s.t.} \ {\mathds 1}_n^{\mathsf T}&[
         k_1^{i}  ,
                \ldots ,
                   k_n^{i} 
]^{\mathsf T}=0,\quad i\in\mathcal I,
\end{align} 
\end{subequations}
 In this case,  for each $i\in\mathcal I$,  let 
 \begin{align}
 z_i&:=[u_1^i ,\ldots,u_{i-1}^i , u_{i+1}^i, \ldots,u_n^i]^\mathsf T,\\
  W_i(w)&:=\left[\!\begin{array}{ccccccc}
               w_1 \!&\! \cdots \!&\! w_{i-1} \!&\!  w_{i+1} \!&\! \cdots \!&\! w_n\\
              1   \!&\!  \cdots  \!&\!  1 \!&\! 1  \!&\! \cdots \!&\! 1
\end{array}\!\right]  .
\end{align}
The sufficient condition  \eqref{PENE1}  of Pareto efficiency reduces to its first-order condition with balanced budget given by \begin{align}  &W_i(w)z_i \label{sqmlq}
=-u_i^i\left[\!\!\begin{array}{cc}
               w_i \\         1\end{array}\!\!\right]
               \!+\!
                \left[\!\!\begin{array}{cc}  \sum\nolimits_{j=1}^n \!w_j\nabla_i f_j(\hat {\bf x}) \\  0
             \end{array}\!\!\right],
             \ i\in\mathcal I,
\end{align}
where $u=[u_1^1,\ldots, u_n^n]^{\mathsf T}$ is designed to satisfy $u=-F(\hat{\mathbf x})$ for implementing a Nash equilibrium at $\hat{\mathbf x}$.  The solvability of  \eqref{sqmlq} is clear because the rank of matrix $W_i(w)$ is 2 and thus  $W_i(w)$ is full row rank. 
In terms of  \eqref{inequa_kk},  it follows from  
 {${\rm null}( {\mathds 1}_n^{\mathsf T})=\{x\in\mathbb R^n:x=[{\mathds 1}_{n-1},-I_{n-1}]^{\mathsf T}z,z\in\mathbb R^{n-1}\}$}
that  the   inequality  \eqref{inequa_kk}
  is  equivalent to $[w,e_i]^{\mathsf T}[ {\mathds 1}_{\textcolor[rgb]{0,0,0}{n-1}},-I_{n-1}]^{\mathsf T}z_i\leq[  \delta_i ,0]^{\mathsf T}$ for $z_i\in\mathbb R^{n-1}$.

Next, let   $\overline w:=[w_1-w_2,\ldots,w_1-w_n]^{\mathsf T},l_1:={\mathds 1}_{n-1}$,  \textcolor[rgb]{0,0,0}{and } $l_i\in\mathbb R^{n-1}$ denote a  unit vector \textcolor[rgb]{0,0,0}{with the element $[l_i]_{i-1}=-1$ for $i\geq 2$}. It follows from  $[ \textcolor[rgb]{0,0,0}{{\mathds 1}_{n-1}},-I_{n-1}][w,e_i]=[\overline w,l_i]$ that  ${\rm rank}\,([ \textcolor[rgb]{0,0,0}{{\mathds 1}_{n-1}},-I_{n-1}][w,e_i])=2$  as no  welfare weight   $w_i$   appears $n-1$
times. Thus, $[ \textcolor[rgb]{0,0,0}{{\mathds 1}_{n-1}},-I_{n-1}][w,e_i]y=0$ in $y\in\mathbb R^2$  has only   a zero solution $y= 0$, which breaks the condition $[  \delta_i ,0] y\not=0$.
Thus, it follows from Farkas' lemma~\cite{matouvsek2007understanding} that $[w,e_i]^{\mathsf T}[ \textcolor[rgb]{0,0,0}{{\mathds 1}_{n-1},-}I_{n-1}]^{\mathsf T}z_i\leq[  \delta_i ,0]^{\mathsf T}$ has a solution in $z_i\in\mathbb R^n$, i.e., 
each linear inequality \eqref{inequa_kk} has a solution. Hence, it follows from Lemma~\ref{lemma:sp1} that the proof for $n\geq 3$ is complete. 

{Now, for $n=2$, from  Lemma~\ref{lemma:sp2},  the maximum social welfare is not larger  than $\mathcal W(\hat{\bf x})$
with $\hat{\bf x}$ defined in \eqref{hatx2} rather than  \eqref{hatx}. Then, we prove  the sufficiency  of  \eqref{hatx2}  to be the optimal target state by showing  the existence of 
 incentive functions. 
 Without loss of generality,  consider $\nabla\!_1\,   {\mathcal W_{\rm u}}(\hat{\bf  x})=0$.  In this case,  by Assumption~\ref{as_social0}, there is      $y_1(x)=-y_2(x)$
such that  $
\nabla\!_i y_i(   \hat{\bf  x})=-\nabla\!_i y_j(   \hat{\bf  x})=-\nabla\!_i f_i(  \hat{\bf  x})$, $i=1,2$, $j\not=i$,
hold with a 
concave $V_2(x)$  and a $V_1(x)$  which is strictly concave in $x_i$ so that $\hat{\bf  x}$ is a Nash equilibrium.   
Now, since    $\nabla\!_1\,    \mathcal X_{\mathcal W_{\rm u}}(\hat{\bf  x})=0$
implies  $\nabla\!_1 y_1( \hat{\bf  x} )=-\nabla\!_1 f_1(\hat{\bf  x} )=\nabla\!_1 f_2(\hat{\bf  x})$,
it holds  \textcolor[rgb]{0,0,0}{that}   
 \begin{align}\nonumber 
 \nabla\!  y_2(\hat{\bf  x})=&[\nabla\!_1  y_2( \hat{\bf  x}),\nabla\!_2  y_2( \hat{\bf  x})]^{\mathsf T}\\ \nonumber
 =&
[-\nabla\!_1  y_1( \hat{\bf  x} ),-\nabla\!_2  f_2( \hat{\bf  x} )]^{\mathsf T}\\  
 =&[-\nabla\!_1  f_2(\hat{\bf  x} ),-\nabla\!_2  f_2( \hat{\bf  x} )]^{\mathsf T}=-\nabla\!  f_2(\hat{\bf  x}).
\end{align}
This implies $\nabla V_2( \hat{\bf  x})=\nabla  f_2(\hat{\bf  x})+\nabla y_2( \hat{\bf  x})=0$  and thus $\hat{\bf  x}$ is the maximum point of function $V_2(x)$. 
Then,  $\hat{\bf  x}$ is  Pareto efficient  in $\mathcal G(V)$.
Now, the proof is complete.
$\hfill\square$

\begin{rem}
\emph{The optimization problem in \eqref{hatx} \textcolor[rgb]{0,0,0}{follows an~intuitive principle, namely, maximizing the   social welfare $\mathcal W(x)$ subject  \textcolor[rgb]{0,0,0}{only} to} the necessary   condition $\hat{\bf x}\in \mathcal X_{\mathcal W_{\rm u}}$ shown in Lemma~\ref{Lema1} for   individual rationality.  
Interestingly, when $n\geq 3$,
\textcolor[rgb]{0,0,0}{this principle is fully implementable for the IRPE incentive mechanisms. In this case, the  design of the regulation   admits sufficient degrees of freedom to simultaneously satisfy the constraints of Nash, Pareto efficiency, and balanced budget.}
In contrast, when $n=2$, these degrees of freedom may be lost. The single inequality constraint in \eqref{hatx} is no longer sufficient to determine a feasible target strategy, and an additional structural condition becomes necessary. As a result, the  incentive  problem becomes over-constrained, and the optimal  target state $\hat{\bf x}$  of 
\eqref{hatx}  may no longer be implementable.
}
\end{rem}

\begin{rem}
\emph{For the weighted social welfare function $\mathcal W(x)=\mathcal W_{\rm w}(x)$,  we note that all the solutions of    optimal target states  \eqref{hatx}  appear on the Pareto front (i.e., the set of Pareto-efficient states) of the system $\mathcal G(f)$ before regulation. To see this, the  Lagrangian $\mathcal L:\mathbb R^n\times \mathbb R_+^0\to\mathbb R$
associated with
 the optimization    \eqref{hatx} is defined with   the  Lagrange  multiplier $\lambda\geq0$ as 
\begin{align} \label{Lagram}
\mathcal L(x,\lambda):= \mathcal W(x)+ \lambda[ {\mathcal W_{\rm u}}(x)- {\mathcal W_{\rm u}}(x_{\rm bs})].
\end{align}
We denote the saddle point of the Lagrangian function
 $\mathcal L(x,\lambda)$ in $x\in\mathbb R^n$ and $\lambda\geq0$ as 
 $(\hat {\bf x},\lambda^*)$. It holds  from \textcolor[rgb]{0,0,0}{Karush-Kuhn-Tucker} (KKT)  theorem \cite{giorgi2016approximate} that 
 \begin{equation} 
\nabla_x \mathcal L(\hat{\bf x},\lambda^*)=\nabla \mathcal W(\hat{\bf x})+\lambda^*\nabla  {\mathcal W_{\rm u}}(\hat{\bf x})={\bf 0}_n.
\end{equation} 
There must be a state $x\in\mathbb R^n$ such that
$\mathcal W_{\rm u}(x)> \mathcal W_{\rm u}(x_{\rm bs})$   for any  $x_{\rm bs}\not=\arg\max_{x\in\mathbb R^n}  \mathcal W_{\rm u}(x)$, and hence
  Slater’s condition holds. 
In this case,   Assumptions~\ref{as_social} and \ref{as_social0}  indicate   strong duality of the primal problem  \eqref{hatx}.
Thus,  $\hat{\bf x}$ is a  solution of    \eqref{hatx} if and only if  $(\hat {\bf x},\lambda^*)$ is a saddle point of the Lagrangian function $\mathcal L(x,\lambda)$ in $x\in\mathbb R^n$, $\lambda\geq0$. Then, it is obtained  that 
  \begin{align} 
\hat{\bf x}=\arg\max_{x\in\mathbb R^n} \sum_{i\in\mathcal I}\omega_if_i(x),\ \ \omega_i:=\xi_i+\lambda^*,\ \ i\in\mathcal I,
\end{align}
and hence $\hat{\bf x}$ is Pareto efficient in $\mathcal G(f)$ due to $\omega_i>0$, $i\in\mathcal I$ (e.g., see    the solution $\hat{\bf x}$ of \eqref{hatx}  illustrated  as  the circular nodes in Fig.~\ref{sasaasa}  under different weights $\xi$ on the grey parabolic curve which represents the Pareto front of $\mathcal G(f)$).}
\end{rem}

\begin{figure}
\centering
\includegraphics[width=0.99\linewidth]{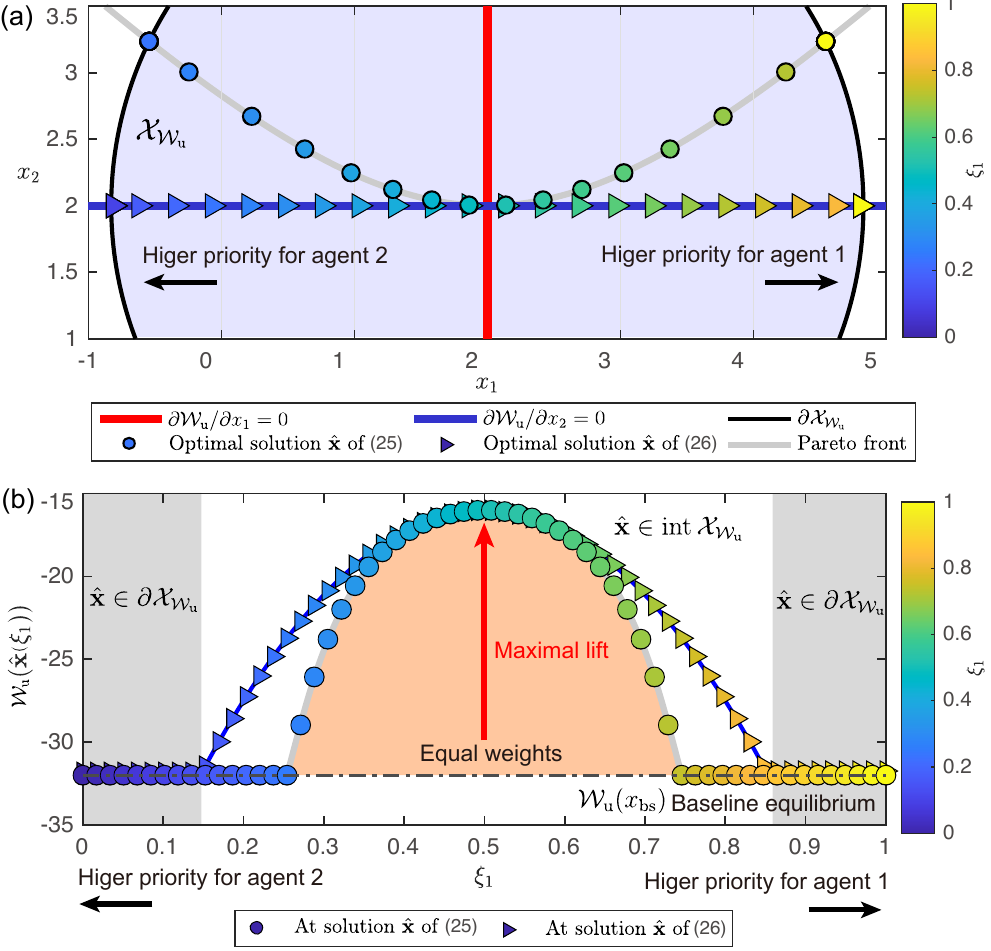}
\caption{Trajectories of the optimal  solutions of \eqref{hatx} and \eqref{hatx2}     and the utilitarian   social welfare $\mathcal W_{\rm u}(\hat{\bf x})$ while   increasing the      weight  $\xi_1\in(0,1)$ for a two-agent game  under a weighted social welfare function $\mathcal W(x)=\mathcal W_{\rm w}(x)$. Note that the blue region in  phase portrait (a) represents the   utilitarian benefiting region $ \mathcal X_{\mathcal W_{\rm u}}$ in Fig.~\ref{siji1}.}
\label{sasaasa}\vspace{-10pt}
\end{figure}

\begin{exa}\label{exa1}\emph{Consider  {the} two-agent noncooperative system with a couple of    quadratic  payoff functions   given by 
$
f_1(x) =-(x_1-4)^2-x_2^2+2x_1x_2$ and $
f_2(x) =-(x_2-4)^2-2x_1x_2$. The Nash equilibrium     of $\mathcal G(f)$ without any incentives is given by  $x^*=[4,0]^{\mathsf T}$. Let the social welfare function $\mathcal W(x)$ be the weighted one in     \eqref{weijt_f} with
    weights          $\xi=[2/3,1/3]^\mathsf T$.  In Fig.~\ref{siji1}, the socially optimal state $\hat x^{\rm opt}$ is illustrated as the blue node, whereas the grey ellipses represent the  level sets of  the social welfare $\mathcal W(x)$. 
Now, let the baseline equilibrium $x_{\rm bs}$ be the   Nash equilibrium     of $\mathcal G(f)$;  the utilitarian benefiting region $\mathcal X_{\mathcal W_{\rm u}}$ in \eqref{XWusa} is depicted as the blue circular region in Fig.~\ref{siji1}. 
In this case, even though $\hat x^{\rm opt}\in\mathcal X_{\mathcal W_{\rm u}}$ holds, it follows from Lemma~\ref{lemma:sp2} that    $\hat x^{\rm opt}$ is an inadmissible target   state as $\hat x^{\rm opt}$ does  not appear on the zero-gradient curves  $\partial {\mathcal W_{\rm u}}(x)/\partial x_i=0$, $i=1,2$, illustrated as the red and  blue straight lines.
The optimal target state $\hat{\mathbf x}$ calculated from \eqref{hatx2} is shown as the red node in Fig.~\ref{siji1}.
Furthermore, Fig.~\ref{sasaasa} illustrates how the welfare  weights  $\xi_1\in[0,1]$ and $\xi_2=1-\xi_1$  influence the optimal target state  $\hat{\bf x}$  and the resulting utilitarian welfare ${\mathcal W_{\rm u}}(\hat{\bf x})$. 
}\end{exa}

In social science, 
a government may assign a larger welfare weight to an agent to promote egalitarianism \cite{o1987egalitarians}.
However,  we introduce an interesting phenomenon that assigning a higher welfare weight to a particular agent   can reduce the total surplus available for redistribution, thereby undermining the possibility of making all agents strictly better off. 
We refer to this phenomenon as  the \emph{agent priority paradox}.
To see this, recall the discussion in Remark~\ref{remal:1} that  the utilitarian surplus ${\mathcal W_{\rm u}}(\hat{\bf x})-{\mathcal W_{\rm u}}(x_{\rm bs})$ determines whether agents can be strictly better off than at the baseline equilibrium,  or they must retain their baseline payoffs with $V_i(\hat{\mathbf{x}})=f_i(x_{\rm bs})$ for $i\in\mathcal I$.
It can be seen from  Fig.~\ref{sasaasa}(b) that  when the welfare weights $\xi$ are nearly identical, this  utilitarian surplus  ${\mathcal W_{\rm u}}(\hat{\bf x})-{\mathcal W_{\rm u}}(x_{\rm bs})$  is strictly positive, which allows all agents to benefit from the incentives.
By contrast, when weight disparities are sufficiently large (see the grey region in Fig.~\ref{sasaasa}(b)),   the optimal target state $\hat{\bf x}$  appears  on   $\partial\mathcal X_{\mathcal W_{\rm u}}$, forcing agents' effective payoffs to return to the  baseline levels.
As a result, prioritizing a single agent increases the likelihood that no agent can be made strictly better off under the mechanism.
Related discussion about the influence of welfare weights $\xi$  will be  further given in Section~\ref{sec:extension}. 

\subsection{Updating  Rules of IRPE Incentive Functions}\label{nonequal_sin}

  In this section, we present the updating rules of   the reference signal $x_{\rm r}(t)$, the gradient vectors  $ u_i(t)$, $ i\in\mathcal I$, (or, equivalently, the control input $u(t)$ and   $z_i(t)$, $ i\in\mathcal I$), and the  scalar function   $\tau_i(x_{\rm r}(t))$ for  the  linear-quadratic incentive functions  \eqref{incen_infla}. Here, we will focus on $n\geq 3$ case since the special case $n=2$ can be treated relatively easily. Notice that for $n=2$, the constrained optimization problem \eqref{hatx2} can be  solved by  offline learning methods. 
In this case, without loss of generality,     given $\hat{\bf x}$ of \eqref{hatx2} satisfying  $\nabla\!_1  \mathcal X_{\mathcal W_{\rm u}}(\hat{\bf  x})=0$,   the Pareto-efficient Nash equilibrium  on $\hat{\bf x}$ can be also easily achieved  using the gradient parameters and the Hessian matrices of \eqref{incen_infla}  given by 
  \begin{align} \nonumber
&u=[u_1^1,u_2^2]^\mathsf T=-F(\hat{\mathbf x}), \ \ u_1^2=-u_2^2, \ \ u_2^1=-u_1^1,\\
&K_2=-K_1=\{K_2\in\mathcal K_2:\nabla^2 f_2(x)+K_2\leq 0\}.
\end{align}
This guarantees a maximum point $\hat{\bf x}$ for the effective function $V_2(x)$ and hence ensures a Pareto-efficient Nash equilibrium at $\hat{\bf x}$.
Therefore, in what follows, we consider only the case~$n\geq 3$ where  the target state $\hat{\bf x}$ in  \eqref{hatx} is not given \emph{a priori}.

Given a baseline  equilibrium $x_{\rm bs}\in\mathbb R^n$,  the following  statements summarize our idea for the linear-quadratic incentives \eqref{incen_infla} to be optimal IRPE.

\subsubsection{Reference Signal for Maximum Attainable Social Welfare} To seek the optimal target state $\hat{\bf x}$ defined in \eqref{hatx}, the reference signal $x_{\rm r}(t)$ for $n\geq 3$ is obtained using the hybrid primal-dual gradient dynamics
\begin{subequations}\label{xrm}
\begin{align}
\dot{x}_{\rm r}(t)=\,&\frac{\partial \mathcal W(x_{\rm r}(t))}{\partial x}+ \lambda_{\rm r}(t) \frac{\partial   {\mathcal W_{\rm u}}(x_{\rm r}(t))}{\partial x},\\ \label{xrm2}
\dot{\lambda}_{\rm r}(t)=\,&\left[ {\mathcal W_{\rm u}}(x_{\rm bs})-  {\mathcal W_{\rm u}}(x_{\rm r}(t)) \right]_{\lambda_{\rm r}}^+,
\end{align} 
\end{subequations}
where 
${\lambda}_{\rm r}\in\mathbb R$ is a scalar. It is clear that the reference signal $x_{\rm r}(t)$ is independent of the  states of the agents as required in Assumption~\ref{as_u_i}. 
 It directly follows from \cite{FEIJER20101974} that  
   $x_{\rm r}(t)\to\hat{\bf x}$ and $\lambda_{\rm r}(t)\to\lambda^*$ as $t\to\infty$  for  any \textcolor[rgb]{0,0,0}{initial conditions} $x_{\rm r}(0)\in\mathbb R^n$,  $\lambda_{\rm r}(0)\in\mathbb R_+$ under Assumptions~\ref{as_social} and \ref{as_social0}.

\subsubsection{Control Input   for Shifting Nash Equilibrium}  Denote $e(t)=x(t)-x_{\rm r}(t)$ as   the  tracking error \textcolor[rgb]{0,0,0}{between} the agents' state $x(t)$  \textcolor[rgb]{0,0,0}{and} the reference signal $x_{\rm r}(t)$. 
Under Assumption~\ref{as_u_i}, to move the Nash equilibrium $\tilde x^*(t)$ of $\mathcal G(V)$, the controller generating $u(t)$ is designed  as 
  \begin{subequations}\label{ut}
 \begin{align} \label{ut1}
u(t)
=&\,{\rm diag}[\exp(\epsilon(t))]^{-1}\dot x_{\rm r}(t)- F( x_{\rm r}(t)),\\ \label{ut2}
\dot {\epsilon}(t)=&\,{\rm diag}[\exp(\epsilon(t))]^{-2}{\rm diag}[\dot x_{\rm r}(t)]e(t),
\end{align}
\end{subequations}
where $\epsilon\in\mathbb R^n$ is an auxiliary state. 
In this case, the  pseudo-gradient dynamics \eqref{dynamics2} of  the incentivized system $\mathcal G(V)$ under  the linear-quadratic incentive functions \eqref{incen_infla} and the updating laws \eqref{xrm}, \eqref{ut}   can be  expressed  as a cascade system given by
   \begin{align} \label{samlq}
\dot x(t) ={\rm diag}[\beta]( F(x(t))+Ke(t)+u(t)),
\end{align}
where $u(t)$ can be understood as the output of the reference system \eqref{xrm}. 
A sufficient condition for convergence of the state $x(t)$ will be discussed  later.

\subsubsection{Guarantee of Pareto Efficiency} As a Nash equilibrium may not be Pareto efficient, the control input $u=[u_1^1,\ldots,u_n^n]^\mathsf T$ alone cannot always ensure Pareto efficiency at  the optimal target state $\hat{\mathbf x}$. Here, we consider how to utilize other gradient  directions  $z_i=[u_1^i ,\ldots,u_{i-1}^i , u_{i+1}^i, \ldots,u_n^i]^\mathsf T$ in $u_i$, $ i\in\mathcal I$.  In particular, 
to guarantee the reference signal $x_{\rm r}(t)$ to  be Pareto efficient  at each  $t\geq 0$ with a balanced budget, we construct the updating rules for  the other directions of gradient vectors  $z_i(t)$, $ i\in\mathcal I$. Specifically, we generate $z_i(t)$  according to  the linear mapping  with respect to the control input $u(t)$ given by 
\begin{align}\label{ztt}
z_i(t)=\kappa_i(w,u(t)), \quad i\in\mathcal I,
\end{align}
with  a common weight $w\in\Delta_+$ and the linear mapping      
\begin{align}\nonumber  
\kappa_i(w,u):=\,&W_i^{\dagger}(w)\left(-u_i^i\!
\left[\!\begin{array}{ccccccc}
               w_i\\
               1
\end{array}\!\right]+ 
\left[\!\begin{array}{ccccccc}
\sum\nolimits_{j=1}^n w_j\nabla_i f_j(x_{\rm r})\\
               0
\end{array}\!\right]\right)\\ \label{ztt00}
&+( {I_{n-1}}\!-\!W_i^{\dagger}(w)W_i(w))\psi,
\end{align} 
where $\psi\in\mathbb R^{n-1}$
and  $(\cdot)^\dagger$ is the Moore-Penrose pseudoinverse.
The sufficient condition on $w$ and the Hessian matrix $K_i$ for the optimal target state $\hat{\bf x}$   to be a Pareto-efficient Nash equilibrium in $\mathcal G(V)$ will be discussed later. 
 
 \subsubsection{Steady-State Payoff Redistribution for Individual Rationality}
The zero-sum scalar function $\tau_i(x_{\rm r})$ determines the amount of payoff transfer $y_i(x_{\rm r})$ at the  reference  state $x_{\rm r}(t)$   to achieve the individual rationality condition 
  \begin{align}\label{IRSM}
\lim_{t\to\infty}V_i(x(t))=\lim_{t\to\infty} f_i(x(t))+\tau_i(x_{\rm r}(t))\geq   f_i( x_{\rm bs}),
  \end{align}
for $i\in\mathcal I$.     To this end, similar to  the necessary and sufficient condition \eqref{offest_eq} in Proposition~\ref{lemma2aq},  the scalar function $\tau_i:\mathbb R^n\to\mathbb R$ is constructed according to the reference signal $x_{\rm r}(t)$ as 
   \begin{align} 
\tau_i(x_{\rm r})=\,&\eta_i[ {\mathcal W_{\rm u}}(x_{\rm r})- {\mathcal W_{\rm u}}(x_{\rm bs})]  \label{ysinkal2}
+f_i(x_{\rm bs})-f_i(x_{\rm r}),
\end{align} 
where     $\eta_i\geq0$, $i\in\mathcal I$, with  $\sum_{i\in\mathcal I}\eta_i=1$.

   \subsubsection{Stability Analysis}In this part, we present our main  
results of stability analysis for the proposed updating rules \eqref{xrm}, \eqref{ut}, \eqref{ztt}, and \eqref{ysinkal2}, where the  sensitivity parameters   $\beta_i$, $i\in\mathcal I$,  in the pseudo-gradient dynamics \eqref{dynamics2}  are unknown. 
The following theorem provides      sufficient conditions to guarantee  global   convergence of the state $x(t)$ towards the optimal target state $\hat {\bf x}$ as a Pareto-efficient Nash equilibrium. Here, let 
\begin{align}  \label{Ajalma}
\mathcal A (\beta,x):={\rm diag}[ \beta]
\left[\begin{array}{ccc}\frac{\partial^2 f_1(x)}{\partial x_1^2} & \cdots & \frac{\partial^2 f_1(x)}{\partial {x_1}\partial{x_n}}\\ \vdots & \ddots & \vdots \\ \frac{\partial^2 f_n(x)}{\partial {x_n}\partial{x_1}} & \cdots & \frac{\partial^2 f_n(x)}{\partial x_n^2}\end{array}\right]
\end{align}
be the Jacobian matrix of the non-incentivized pseudo-gradient dynamics $\dot x(t) ={\rm diag}[\beta]F(x(t))$ at the state $x\in\mathbb R^n$.
Then, 
the Jacobian matrix of the pseudo-gradient dynamics \eqref{samlq} for $u(t)=u^*$ and $x_{\rm r}(t)=\hat{\bf x}$ is given by 
$\Pi (\beta,x):=\mathcal A(\beta,x)+{\rm diag}[ \beta]K\in\mathbb R^{n\times n}$.

  \begin{thm} \label{smalX}
Consider the incentivized noncooperative system $\mathcal{G}(V)$  under  the pseudo-gradient dynamics \eqref{dynamics2} and Assumptions~\textcolor[rgb]{0,0,0}{\ref{as_Ji}, \ref{as_social}, and \ref{as_social0}} with $n\geq 3$ agents.
Let the incentive functions $y_i(x)$, $i\in\mathcal I$, be given by \eqref{incen_infla}.
If the matrices ${ K}_i\in\mathcal K_i$, $ i\in\mathcal I$,  satisfy $K_1+\cdots+K_n={\bf 0}_{n\times n}$ and  guarantee the existence of  $w=[w_1,\ldots,w_n]^\mathsf T\in\Delta_+^{lr}$,   $\hat\beta\in\mathbb R_+^n$, and $\zeta\in\mathbb R_+$   such that
  \begin{subequations}\label{condition_thm}
  \begin{align}\label{condition2}
   \sum\nolimits_{i\in\mathcal I}w_i\Big[\nabla^2 f_i( x)+K_i\Big]&\leq0,
\\
\label{condition1}
{\Pi^\mathsf T (\hat\beta,x)+\Pi( \hat\beta,x)}&<-\zeta I_n,\ \, x\in\mathbb R^n,
\end{align}
\end{subequations}
then the updating laws of    \eqref{xrm}, \eqref{ut}, \eqref{ztt}, and \eqref{ysinkal2}   guarantee that the trajectory of agents' state $x(t)$ globally  converges  to the optimal target state $\hat{\bf x}$  defined in \eqref{hatx}     as a Pareto-efficient Nash equilibrium with a strong balanced budget \eqref{bdgt} and the individual rationality \eqref{IRSM} for any  sensitivity parameters $\beta_i\in\mathbb R_+$, $i\in\mathcal I$,  initial state $x(0)\in\mathbb R^n$,   free parameter vector  $\psi\in {\mathbb R^{n-1}}$, auxiliary state $\epsilon(0)\in\mathbb R^n$, and initial values $x_{\rm r}(0)\in\mathbb R^n$ and $\lambda_{\rm r}(0)\in\mathbb R_+$. 
 \end{thm}

  \noindent\textbf{Proof} First, notice that the function   $\sum\nolimits_{i\in\mathcal I}  w_i V_i(x)$ is concave under the condition \eqref{condition2}, whereas \eqref{ztt} solves the first-order condition of  \eqref{PENE1}. Hence, it follows from Lemma~\ref{lemma:sp1} that $x_{\rm r}(t)$ is  Pareto efficient in $\mathcal G(V)$ for $t\geq0$.
Note from \eqref{samlq} that the tracking error dynamics of $
e(t)=x(t)-x_{\rm r}(t) 
$ are given by 
  \begin{align}\nonumber 
\dot e(t)=\,&{\rm diag}[\beta](F(x(t))+Ke(t)+u(t))-\dot x_{\rm r}(t)\\ \nonumber
=\,&{\rm diag}[\beta](F(x(t))-F(x_{\rm r}(t))+Ke(t))\\  \label{dyna_et}
&+{\rm diag}\left[(\beta-\exp(\epsilon(t)))\oslash \exp(\epsilon(t))\right]\dot x_{\rm r}(t),
\end{align}
whereas the dynamics of the estimation error $\tilde\beta:=\beta-\check\beta$ with $\check\beta:= \exp(\epsilon)$ are given from \eqref{ut2} by 
  \begin{align} \label{dyna_ebeta}
\dot {\check\beta}(t)=-{\rm diag}[\check\beta(t)]^{-1}{\rm diag}[\dot x_{\rm r}(t)]e(t).
\end{align}
Now, let $P:={\rm diag}\big[ \hat\beta \oslash \beta\big]>0$ and 
\begin{equation}
\sigma:=\left\{\begin{array}{ll}
0, &{\rm if}\,\lambda_{\rm r}=0\ {\rm and}\, \mathcal W_{\rm u}(x_{\rm r})- \mathcal W_{\rm u}(x_{\rm bs})>0,\\
1, &{\rm otherwise}.
\end{array} \right.
\end{equation}
Note that $\dot{\lambda}_{\rm r}(t)=0$ for $\sigma(t)=0$. Consider the   function 
\begin{align} \label{Lafu0} 
V(e , x_{\rm r},\lambda_{\rm r},\tilde\beta,\sigma)\!&:=\frac{1}{2}\big[e^{\mathsf T}Pe+\tilde \beta^{\mathsf T}P\tilde\beta \big]\!+\!V_0(x_{\rm r},\lambda_{\rm r},\sigma),
 \\
 \label{Lafu}
V_0(x_{\rm r},\lambda_{\rm r},\sigma)\!&:=\frac{1}{2}\dot x_{\rm r}^{\mathsf T}\dot x_{\rm r}+\frac{1}{2}|\dot\lambda_{\rm r}|^2 \sigma,
\end{align}
for the entire  system with the reference dynamics   \eqref{xrm} and the   error dynamics \eqref{dyna_et}, \eqref{dyna_ebeta}.
Note that the condition \eqref{condition1} implies the Jacobian matrix $\Pi( \hat\beta,x)$ uniformly~negative definite in $\mathbb R^n$ as $x^{\mathsf T  }\Pi(\hat\beta,x)x<-\zeta\|x\|^2$. 
It follows from the property of monotone vector mappings that the vector 
\begin{align}
g(x,\hat\beta):={\rm diag}[\hat\beta](F(x)+Kx)
\end{align}
is strongly monotone in $\mathbb R^n$, i.e., there exists $\zeta_0$ such that 
$(x-y)^{\mathsf T  }(g(x,\hat\beta)-g(y,\hat\beta))\leq-\zeta_0\|x-y\|^2$
holds for any $x,y\in\mathbb R^n$.
 Then,  for a time interval
with fixed  $\sigma$,
 it holds from \eqref{condition1}  that 
 \begin{align}\nonumber 
&\dot V(e(t) , x_{\rm r}(t),\lambda_{\rm r}(t),\tilde\beta(t),\sigma)\\ \nonumber
&=e^{\mathsf T}\!(t)({\rm diag}[\hat\beta](F(x(t))\!- \!F(x_{\rm r}(t))   \! +\!Ke(t))\!\\ \nonumber
&\ \quad\quad+\!{\rm diag}[\tilde\beta(t)\!\oslash\! \check\beta(t)\! \odot\! \hat\beta \!\oslash \!\beta ]\dot x_{\rm r}(t)\!+\!\dot V_0(x_{\rm r}(t),\lambda_{\rm r}(t),\sigma ) \\  \nonumber
&\ \quad\quad-\tilde\beta^{\mathsf T}(t){\rm diag}\big[ \hat\beta \oslash \beta\big]{\rm diag}[\check\beta(t)]^{-1}{\rm diag}[\dot x_{\rm r}(t)]e(t)
\\ 
&\leq-\zeta_0\|e(t)\|^2+ \dot V_0(x_{\rm r}(t),\lambda_{\rm r}(t),\sigma ).
\end{align}
\textcolor[rgb]{0,0,0}{Also, it holds from Assumptions \ref{as_social} and \ref{as_social0} that}
\begin{align}\nonumber 
&\dot V_0(x_{\rm r}(t),\lambda_{\rm r}(t),\sigma )=\dot x_{\rm r}^{\mathsf T}(t)\ddot x_{\rm r}(t)+\dot\lambda_{\rm r}(t)\ddot\lambda_{\rm r}(t) \sigma \\ \nonumber
&=\dot x_{\rm r}^{\mathsf T}(t)\left[
\nabla^2 \mathcal W(x_{\rm r}(t))+ \lambda_{\rm r}(t) \nabla^2   \mathcal W_{\rm u}(x_{\rm r}(t)) \right]\dot x_{\rm r}(t)\\  \nonumber
&\quad+\dot x_{\rm r}^{\mathsf T}(t)\nabla  \mathcal W_{\rm u}(x_{\rm r}(t))\dot\lambda_{\rm r}(t)
\!-\!\dot\lambda_{\rm r}(t) \nabla  \mathcal W_{\rm u}(x_{\rm r}(t)) ^{\mathsf T}\dot x_{\rm r}(t) \sigma
\\ \label{eamnlsa}
&= \dot x_{\rm r}^{\mathsf T}(t)\!\left[
\nabla^2 \mathcal W(x_{\rm r}(t))\!+\! \lambda_{\rm r}(t) \nabla^2   \mathcal W_{\rm u}(x_{\rm r}(t)) \right]\!\dot x_{\rm r}(t)\leq 0. 
\end{align}

During the switching of $\sigma$, if $\sigma$ turns from 0 to 1 at $t_0^+$, then even though the second term in \eqref{Lafu} becomes nonzero, there is no discontinuity
of $V_0(x_{\rm r},\lambda_{\rm r},\sigma)$  and $V(e , x_{\rm r},\lambda_{\rm r},\tilde\beta,\sigma)$ since  $\dot{\lambda}_{\rm r}(t_0)=0$. Next, consider when $\sigma$ turns from 1  to 0 at time $t_0^+$, i.e.,  $ \mathcal W_{\rm u}(x_{\rm r}(t_0))- \mathcal W_{\rm u}(x_{\rm bs})>0$,    $\lambda_{\rm r}(t_0)>0$,  and $ {\lambda_{\rm r}}(t_0^+)=0$. In this case, as  $V_0(x_{\rm r},\lambda_{\rm r},\sigma)$  \textcolor[rgb]{0,0,0}{loses}
 a positive term, $V_0(x_{\rm r},\lambda_{\rm r},\sigma)$ is  discontinuous   in a decreasing direction. 
Thus,   the  function $V(e , x_{\rm r},\lambda_{\rm r},\tilde\beta,\sigma)$ is  nonincreasing along  the trajectories of the closed-loop dynamics  \eqref{xrm}, \eqref{dyna_et}, and \eqref{dyna_ebeta}.

Next, we must show      from the LaSalle invariance principle for hybrid systems  \cite{lygeros2003dynamical,FEIJER20101974}  that every trajectory $(x_{\rm r},\lambda_{\rm r})$ of \eqref{xrm} in $\mathbb R^n\times\mathbb R_+$ converges to
 the maximal positively invariant set satisfying:
 
 1) $\dot V_0(x_{\rm r}(t),\lambda_{\rm}(t),\sigma)\equiv 0$ at every interval of fixed $\sigma$;
 
 2)  $\dot V_0(x_{\rm r}(t),\lambda_{\rm}(t),\sigma^-)=\dot V(x_{\rm r}(t),\lambda_{\rm}(t),\sigma^+)$ if $\sigma$ switches at $t$ from $\sigma^-$ to $\sigma^+$. 

We show that the maximal positively invariant set  satisfying 1) and 2) must be an equilibrium   $(\hat {\bf x},\lambda^*)$ using the   function \eqref{Lafu}.
We note \textcolor[rgb]{0,0,0}{from} \eqref{eamnlsa} that 
 $\dot V_0(x_{\rm r}(t),\lambda_{\rm}(t),\sigma)\equiv 0$ requires $\dot x_{\rm r}(t)\equiv 0$ under Assumptions \ref{as_social} and \ref{as_social0}, i.e., $x_{\rm r}(t)\equiv\hat {\bf x}$. 
Then, we consider three cases. First, when $ \mathcal W_{\rm u}(\hat {\bf x})- \mathcal W_{\rm u}(x_{\rm bs})<0$,
 from
\eqref{xrm}, the  multiplier $\lambda_{\rm r}(t)$ grows unboundedly, which contradicts the   
boundedness result \cite[Proposition 3]{FEIJER20101974}. 
Second, when $\mathcal W_{\rm u}(\hat{\bf x})- \mathcal W_{\rm u}(x_{\rm bs})>0$,  note that  $\lambda_{\rm r}(t)\equiv0$ must hold. Otherwise, it follows from \eqref{xrm} that $\lambda_{\rm r}(t)$  converges linearly to zero, which results  in a discontinuity of $V_0(x_{\rm r}(t),\lambda_{\rm}(t),\sigma)$ and hence contradicts with  2). 
 Third, when $\mathcal W_{\rm u}(\hat{\bf x})- \mathcal W_{\rm u}(x_{\rm bs})=0$, it follows that $\dot \lambda_{\rm r}(t)\equiv0$. 
Hence, $\dot \lambda_{\rm r}(t)\equiv0$ and $ \lambda_{\rm r}(t)\equiv\lambda^*$.  As a result, we conclude  that the  maximal positively invariant set is  the    saddle
point $(\hat{\bf x},\lambda^*)$ of the Lagrangian. Thus, it follows from the LaSalle invariance principle for hybrid systems  that   $x_{\rm r}(t)\to\hat{\bf x}$, $\lambda_{\rm r}(t)\to\lambda^*$  for  any \textcolor[rgb]{0,0,0}{initial conditions} $x_{\rm r}(0)\in\mathbb R^n$,  $\lambda_{\rm r}(0)\in\mathbb R_+$.
Similarly, we can obtain  $e(t)\to{\mathbf 0}_n$ for the whole system, and hence $x(t)$ converges to $\hat{\bf x}$ as a Pareto-efficient Nash equilibrium since the control input \eqref{ut1} implies \eqref{ui_cond}.

The individual rationality  \eqref{IRSM} is satisfied under the convergence of $x(t)$ and $x_{\rm r}(t)$ by  \eqref{ysinkal2} since  \begin{align} \nonumber
\lim_{t\to\infty}V_i(x(t))&=  f_i(\hat{\bf x})+\tau_i(\hat{\bf x})\\
&=\eta_i[ {\mathcal W_{\rm u}}(\hat{\bf x})- {\mathcal W_{\rm u}}(x_{\rm bs})]+f_i(x_{\rm bs}),
  \end{align}
  with  $\hat{\mathbf{x}}\in\mathcal X_{\mathcal W_{\rm u}}$, i.e., ${\mathcal W_{\rm u}}(\hat{\bf x})- {\mathcal W_{\rm u}}(x_{\rm bs})\geq0$.
Furthermore, the strong balanced budget \eqref{bdgt} is satisfied by  \eqref{ztt} and \eqref{ysinkal2} since \eqref{balanced_KU}  holds. The proof is complete.   $\hfill\square$  

 {We   emphasize that  Theorem~\ref{smalX} provides a systematic method to construct incentive mechanisms   that align social efficiency with individual benefits without inflicting losses on any agents. To the best of our knowledge, this is the first work to study such settings for general concave games (cf., e.g., \cite{riehl2018incentive,colini2016approximately,fabiani2023incentives,ferguson2021effectiveness,ratliff2020adaptive,yan2021hierarchical,brown2020can,yan2021Stabilization,yan2023pareto}).}
 
 \begin{rem}
\emph{To determine the suitable   Hessian matrices $K_i={\rm diag}[k_i^1,\ldots,k_i^n]$, $i\in\mathcal I$,   satisfying  \eqref{condition_thm} and $K_1+\cdots+K_n={\bf 0}_{n\times n}$ in Theorem~\ref{smalX} for the linear-quadratic incentive functions \eqref{incen_infla},  additionally under  Assumption~\ref{as_Ji2}, it is always possible to  
address the linear inequality 
 \begin{subequations}\label{inequa_kknew}
 \begin{align}
[w,e_i]^{\mathsf T}&[   k_1^{i} ,  \ldots ,   k_n^i
]^{\mathsf T}\leq[  \delta_i ,\bar\delta_i]^{\mathsf T},
\\
{\rm s.t.} \ {\mathds 1}_n^{\mathsf T}&[
         k_1^{i}  ,
                \ldots ,
                   k_n^{i} 
]^{\mathsf T}=0,\quad i\in\mathcal I,
\end{align} 
\end{subequations}
for $w\in\Delta_+^{lr}$,    where $\delta_i$ is defined in \eqref{def:taui} and  $\bar\delta_i:=\min\{ 0,\delta_{ii}\}$ with $\delta_{ii}$   defined in  \eqref{def:tauis}, $i\in\mathcal I$. Under  Assumption~\ref{as_Ji2},   note that  $\delta_{ii}$, $\delta_i$,  and $\bar\delta_i$ are finite  numbers.  The feasibility of \eqref{inequa_kknew} can be seen from Farkas’ lemma \cite{matouvsek2007understanding} as similarly handled for \eqref{inequa_kk} in the proof of Theorem~\ref{thm:main1} for each $w\in\Delta_+^{lr}$. Hence, there    are infinitely many optimal IRPE incentive functions   yielding   equilibrium outcomes at the target state   $\hat{\bf x}$. 
}
\end{rem}

\begin{rem}
\emph{The condition \eqref{condition1} ensures a unique Nash equilibrium $\tilde x^*(t)$ for   $t\geq0$  in $\mathcal G(V)$.  Suppose, \emph{ad absurdum}, at time $t$, there exists $\hat x^*\in\mathbb R^n$ satisfying $h(\hat x^*,\beta)={\bf 0}_n$ and $\tilde x^*\not=\hat x^*$, where $h(x,\beta):={\rm diag}[\beta](F(x)+Kx-Kx_{\rm r}+u)$.
In this case, letting $x(\theta)=\theta \tilde x^*+(1-\theta)\hat x^*\in\mathbb R^n$ for $\theta\in[0,1]$, 
it follows from 
$\frac{{\rm d} Ph(x(\theta),\beta)}{{\rm d} \theta}=\frac{\partial  Ph(x,\beta)}{\partial x}\frac{{\rm d} x(\theta)}{{\rm d} \theta}=\Pi({\hat\beta},x(\theta))(\tilde x^*-\hat x^*)$ and 
$\int_0^1\frac{{\rm d} Pg(x(\theta),\beta)}{{\rm d} \theta}{\rm d} \theta={\mathbf 0}_n$ with $\hat x=\tilde x^*- \hat  x^*$ that
\begin{align}\nonumber 
{\bf 0}_n&=\hat x^{\mathsf T  }\int_0^1 \frac{{\rm d} (Pf(x(\theta),\beta))}{{\rm d} \theta}{\rm d} \theta
\\ \nonumber 
& =\hat x^{\mathsf T  }\left[\int_0^1 (\Pi^\mathsf T ({\hat\beta},x(\theta))+\Pi({\hat\beta},x(\theta)) ){\rm d}\theta\right]\!\hat x
\end{align}
implies 
$\hat x={\bf 0}_n$, i.e., $\hat x^*=\tilde x^*$, and hence $\tilde x^*$ is a unique Nash equilibrium at time $t$.
}
\end{rem}

The next corollary  shows  sufficient conditions for the special case of   incentive functions in the linear form  to guarantee  global   convergence of the state $x(t)$ towards the optimal target state $\hat {\bf x}$ as a Pareto-efficient Nash equilibrium.

  \begin{cor} \label{cor11}
Consider the incentivized noncooperative system $\mathcal{G}(V)$  under  the pseudo-gradient dynamics \eqref{dynamics2} and Assumptions~\textcolor[rgb]{0,0,0}{\ref{as_Ji}, \ref{as_social}, and \ref{as_social0}} with $n\geq 3$ agents. 
Let the incentive functions $y_i(x)$, $i\in\mathcal I$, be given by \eqref{incen_infla} with $K_i={\bf 0}_{n\times n}$, $i\in\mathcal I$.
If   the social welfare function $\mathcal W(x)$ is given by the weighted social  welfare \eqref{weijt_f} with  $\xi\in\Delta_+^{lr}$,  and there exist  $\hat\beta\in\mathbb R_+^n$, $\zeta\in\mathbb R_+$  such that
  \begin{align}\label{condition12}
{\mathcal A^\mathsf T (\hat\beta,x)+\mathcal A( \hat\beta,x)}&<-\zeta I_n,\ \, x\in\mathbb R^n,
\end{align}then the updating laws of    \eqref{xrm}, \eqref{ut},   \eqref{ysinkal2},   and 
\begin{equation}\label{ztt2}
z_i(t)=\kappa_i^{\rm re}(\xi+\lambda_{\rm r}(t),u(t)), \quad i\in\mathcal I,
\end{equation}
\begin{equation} \nonumber
\kappa_i^{\rm re}(w,u):=-u_i^iW_i^{\dagger}(w)
\left[\!\!\begin{array}{ccccccc}
               w_i\\
               1
\end{array}\!\!\right]\!+\!( {I_{n-1}}\!-\!W_i^{\dagger}(w)W_i(w))\psi
\end{equation} 
guarantee that the trajectory of agents’ state $x(t)$ globally  converges  towards the optimal target state $\hat{\bf x}$  defined in \eqref{hatx}  as a Pareto-efficient Nash equilibrium with a strong balanced budget \eqref{bdgt} and the individual rationality \eqref{IRSM} for any  sensitivity parameters $\beta_i\in\mathbb R_+$, $i\in\mathcal I$,  initial state $x(0)\in\mathbb R^n$,     free parameter vector  $\psi\in {\mathbb R^{n-1}}$, auxiliary state $\epsilon(0)\in\mathbb R^n$,    initial values $x_{\rm r}(0)\in\mathbb R^n$,  and $\lambda_{\rm r}(0)\in\mathbb R_+$. 
 \end{cor}
 
   \noindent\textbf{Proof} Note that the condition \eqref{condition2} is satisfied under Assumption~\ref{as_social} for $w=\xi+\lambda_{\rm r}(t)$, which converges to $ \xi+\lambda_{\rm r}^*$ satisfying $\sum\nolimits_{j=1}^n (\xi_j+\lambda_{\rm r}^*)\nabla_i f_j(\hat{\bf x})=\mathbf 0_n$. Therefore, \eqref{ztt2} reduces to the same value as $t\to\infty$ as \eqref{ztt}. Since \eqref{condition1} is satisfied by \eqref{condition12}, the result is a direct consequence of Theorem~\ref{smalX}.   $\hfill\square$  

 \begin{rem}
\emph{Corollary~\ref{cor11} implies that for weighted social welfare function $\mathcal W(x)=\mathcal W_{\rm w}(x)$, the updating laws of   IRPE incentive mechanisms can be decoupled from the gradient functions $\nabla_i f_j(x)$, $j\not=i$, which are originally  required for \eqref{ztt00} in Theorem~\ref{smalX}.  
}
\end{rem}

\begin{rem} \label{rem10}
\emph{
\textcolor[rgb]{0,0,0}{The noncooperative system  $\mathcal G(f)$ satisfying the condition \eqref{condition12}  is known as  a strongly monotone game where the Nash equilibrium is always unique. Under linear incentive functions, note that the uniqueness of Nash equilibrium remains in the incentivized noncooperative system $\mathcal G(V)$. Typical examples can be found in oligopoly markets discussed in Section~\ref{sec:extension} below.}
}
\end{rem}

\section{Applications  to  Economic Systems}\label{sec:extension}
In this section,  we present three numerical examples from   economic systems   with strategic complements  \cite{JACKSON201595}  and substitutes   \cite{haraguchi2016cournot} to demonstrate the utility of the proposed approach. 

\subsubsection{Application to  Networked Game of R\&D Partnerships}
Consider a  networked game $\mathcal G(f)$ with the underlying strategic complements of R\&D partnerships among $n$ firms represented by an undirected graph, where  \textcolor[rgb]{0,0,0}{each node is a firm (agent)} and each edge represents  a joint development agreement between two firms.
Connections  in the graph reduce the firm $i$'s    costs for producing   $x_i\in\mathbb R_+$  quantity of products  given by \cite{JACKSON201595} 
\begin{equation}\label{sqmlnlkwq}
c_i(x)=\phi_0-\phi\sum\nolimits_{j\in\mathcal I}[A]_{ij}x_j,\quad i\in\mathcal I.
\end{equation}
Here, $\phi_0\in\mathbb R_+$ is the marginal cost without collaboration, $\phi\in\mathbb R_+$ is the intensity of cost reduction induced by neighboring firms, and  $A\in\mathbb R^{n\times n}$ denotes the adjacency matrix with $[A]_{ij}\in\{0,1\}$ representing the absence or presence   of  joint development between  $i$  and  $j$.
Note that \eqref{sqmlnlkwq} reflects the fact that firms' marginal costs decrease  by   collaborating  with more partners.
Given  the cap price $\theta\in\mathbb R_+$, using an inverse market demand function $p_i(x)=\theta-\frac{1}{2}x_i$,  we consider Cournot competition  where the firms compete in quantities  according to
 the     payoff functions   \cite{JACKSON201595}  given by 
\begin{align}\nonumber
f_i(x)&=x_i[p_i(x)-c_i(x)]\\ \label{netws}
&=-\frac{1}{2}x_i^2+x_i\Big(\phi\sum_{j\not=i} [A]_{ij}x_j+\alpha\Big),
\end{align} 
where $\alpha=\theta-\phi_0$ is  the marginal benefit. 
The pseudo-gradient function of   $\mathcal G(f)$ is given by $F(x)=(\phi A-I)x+\alpha 1_n$, where the Jacobian matrix  is negative definite if  $\phi\rho(A)\!<\!1$. Note that the Nash equilibrium without regulation is given by~$x^*=\alpha(I-\phi A)^{-1} 1_n$. 
Moreover, Assumption~\ref{as_social0} holds when $2\phi\rho(A)<1$. 
An 86-node  unweighted  collaboration network obtained  from \cite{nr}  is used in this example~(see~Fig.~\ref{san_net1}). 

\begin{figure}
\centering
\includegraphics[width=0.82\linewidth]{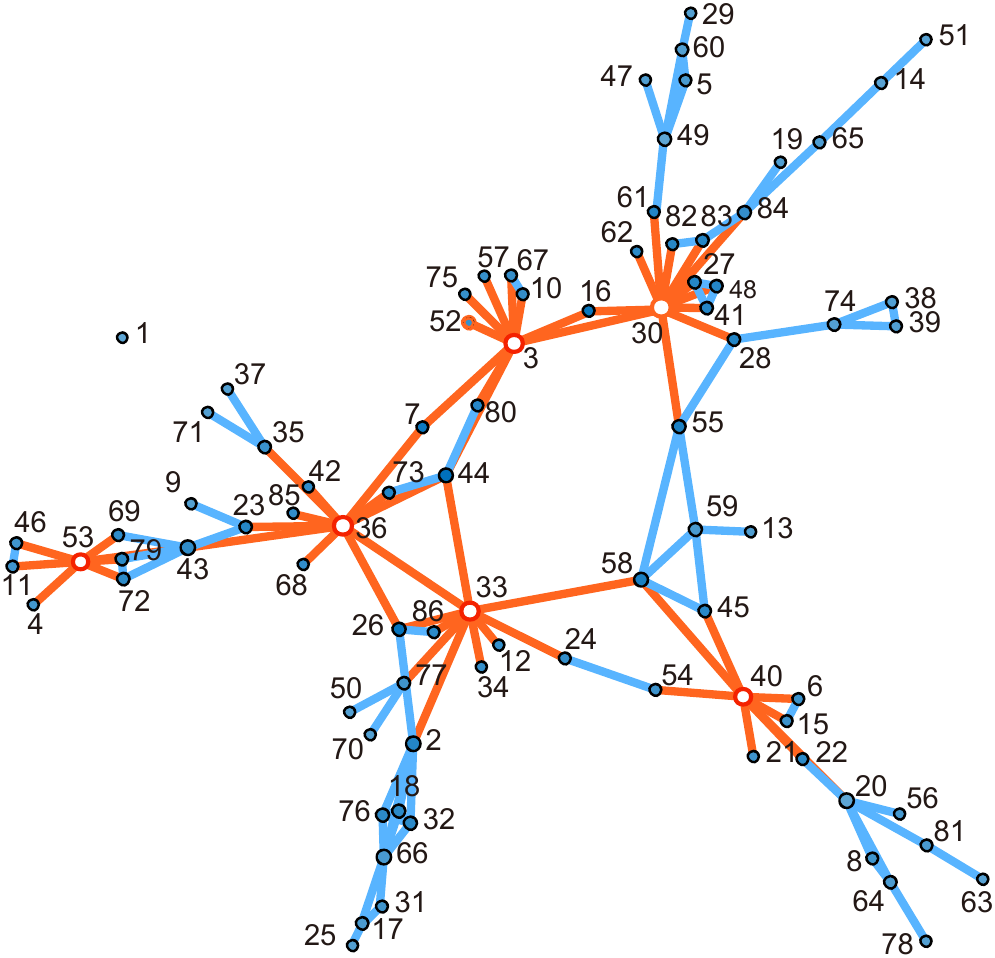}
\caption{{A collaboration network of R\&D partnerships among 86 agents (firms)  \cite{nr}) for a networked game $\mathcal G(f)$}. 
}
\label{san_net1}\vspace{-10pt}
\end{figure}

\begin{figure}[t!]
\centering
\includegraphics[width=0.99\linewidth]{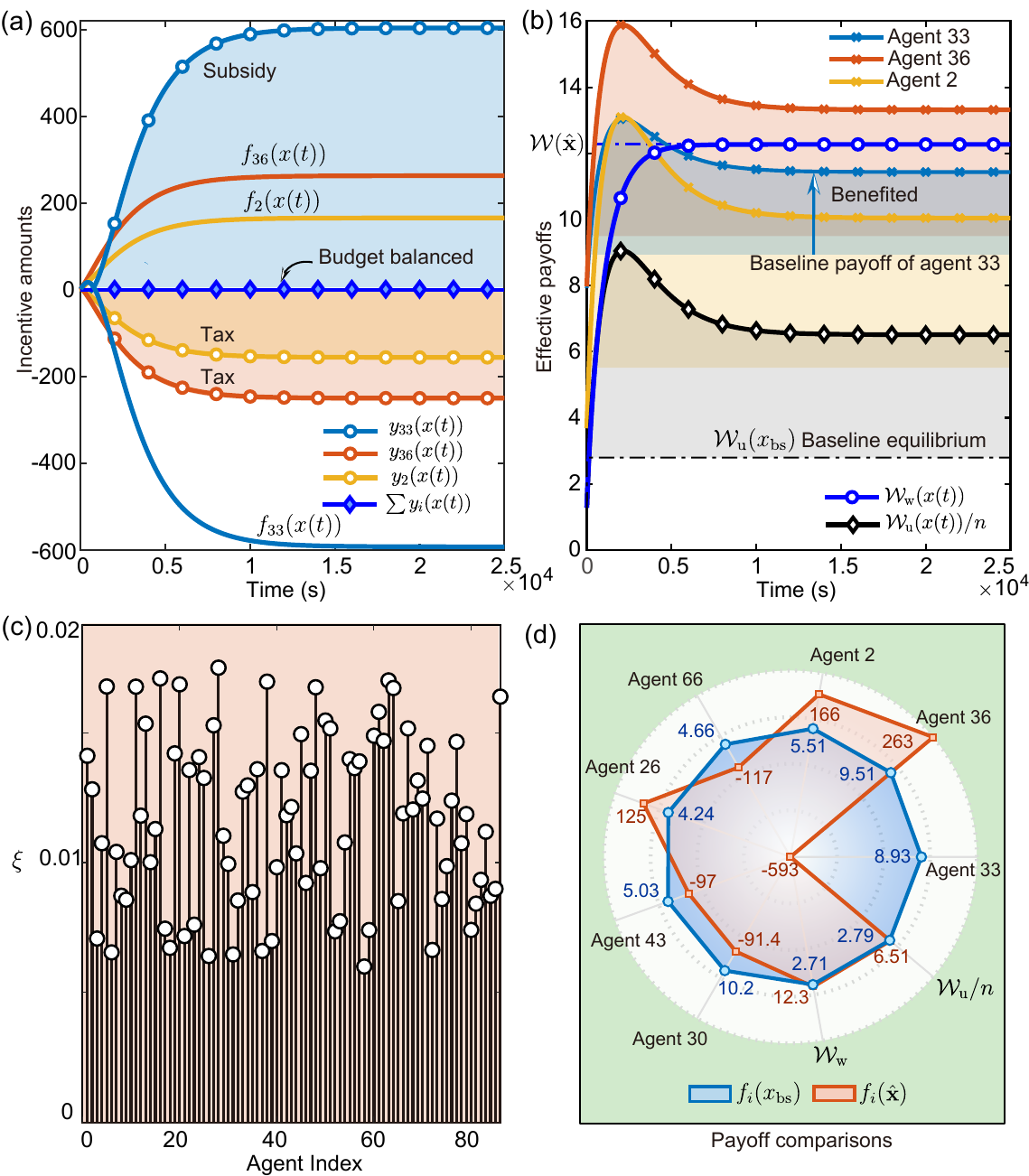}
\caption{The performance of the proposed  incentive mechanism    for an 86-agent   networked game under the  weighted  social welfare function.
 (a)  Trajectories of the agents'  incentives $y_i(x(t))$ and  payoffs  $f_i(x(t))$. (b)  Trajectories of agents' effective payoffs $V_i(x(t))$,  utilitarian social welfare $\mathcal W_{\rm u}(x(t))$, and  weighted social  welfare $\mathcal W_{\rm u}(x(t))$. (c) Weights  $\xi$ in weighted social welfare.   (d)    Payoff levels $f_i(x_{\rm bs})$ and $f_i(\hat{\bf x})$ at the baseline equilibrium $x_{\rm bs}$ and the optimal target state $\hat{\bf x}$.
The total sum of incentive values of all 86 agents is zero for all time $t$ in (a), thereby indicating the balanced budget. 
}\vspace{-10pt}
\label{san_net2}
\end{figure}

\subsubsection*{Weighted social welfare} We   first apply the proposed mechanism for this networked  game  under  weighted social welfare function \eqref{weijt_f}, i.e.,   $\mathcal W(x)=\mathcal W_{\rm w}(x)$, with  randomly  generated   weights   $\xi\in\Delta_+^{lr}$ satisfying Assumption~\ref{as_social}   (see~$\xi$    in  Fig.~\ref{san_net2}(c)).  The baseline equilibrium $x_{\rm bs}$ is set to the Nash equilibrium without regulation, i.e.,  $x_{\rm bs}=x^*$.  
Now, let $\phi=0.1$, $\alpha=1.5$. It follows from $2\phi\rho(A)<1$ that Assumptions~\ref{as_Ji} and \ref{as_social0}  are also satisfied. Note that $\hat x^{\rm opt}\in{\rm int}\mathcal X_{\mathcal W_{\rm u}}$ and hence by Theorem~\ref{thm:main1}, the optimal target state $\hat{\bf x}$ is exactly $\hat x^{\rm opt}$.
Since \eqref{condition12} holds for $\hat \beta={\mathds 1}_n$,  from Corollary~\ref{cor11}, the  updating laws of    \eqref{xrm}, \eqref{ut},   \eqref{ysinkal2},   and \eqref{ztt2} guarantee  that the trajectories of the state $x(t)$ under   the  linear  incentive functions  \eqref{incen_infla} with $K_i={\mathbf 0}_{n\times n}$, $i\in\mathcal I$,  globally   converge  towards the optimal target state $\hat{\bf x}=\hat x^{\rm opt}$  as a Pareto-efficient  Nash equilibrium with balanced budget and  individual rationality. 

In the simulation,  
we let $x(0)=x_{\rm r}(0)=\epsilon(0)=\mathds 1_n$    and $\lambda(0)=10^{-9}$. Furthermore, let  $\eta_i=\xi_i$, $i\in\mathcal I$ for \eqref{ysinkal2} so that  each agent at the optimal target state $\hat{\bf x}$ can additionally  receive $\xi_i[ {\mathcal W_{\rm u}}(\hat{\bf x})- {\mathcal W_{\rm u}}(x_{\rm bs})]$ payoffs compared to the baseline payoff $f_i(x_{\rm bs})$.  
Fig.~\ref{san_net2}(a) illustrates the trajectories of  agents' payoff values $f_i(x(t))$ and the incentives $y_i(x(t))$ for $i=2,33,36$, whereas Fig.~\ref{san_net2}(b) shows the  trajectories of the effective payoffs $V_i(x(t))$, the averaged utilitarian social welfare $\mathcal W_{\rm u}(x(t))/n$, and the social welfare  $\mathcal W_{\rm w}(x(t))$.
Moreover,
 the payoff levels $f_i(x_{\rm bs})$ and $f_i(\hat{\bf x})$ at the baseline equilibrium $x_{\rm bs}$ and the optimal target state $\hat{\bf x}$ are illustrated in Fig.~\ref{san_net2}(d) for agents $i=2,26,30,33,36,43,66$. 
We observe  that  the trajectory of the social welfare $\mathcal W_{\rm w}(x(t))$   converges to the optimal target level $\mathcal W(\hat{\bf x})$ in Fig.~\ref{san_net2}(b) with a balanced budget in Fig.~\ref{san_net2}(a). 
Moreover, some individual payoffs $f_i(x(t))$   might be degraded at the  steady   state (e.g., see the typical nodes 33 and 66), but they are compensated by the subsidies    collected  from  other agents as taxes. 
As a result, in Fig.~\ref{san_net2}(b), the effective payoffs $V_i(x(t))$ of all agents  are substantially improved  compared to the case  without incentives, indicating that the   mechanism is individually rational  at steady state $\hat{\bf x}$  and thereby verifying the results of  Corollary~\ref{cor11}.

\subsubsection*{Discussion on whom should be subsidized for utilitarian social welfare} To determine the necessity to compensate (subsidize) agents' payoff levels for individual rationality, it  is important to discuss the sign of  $\Delta_i:=f_i(\hat{\bf x})-f_i(x_{\rm bs})$. Here,  note that we must give subsidies $ \tau_i(\hat{\bf x})>0$ to    agent $i$ with  $\Delta_i<0$ from   taxing   another  agent $j$ with $\Delta_j>0$.  In a special case~when  $\Delta_i>0$, $i\in\mathcal I$, it is clear that the steady-state transfers $y_i(\hat{\bf x})= \tau_i(\hat{\bf x})$, $i\in\mathcal I$, of the incentive functions \eqref{incen_infla} are not necessary. 
To examine the condition of zero steady-state transfers  $y_i(\hat{\bf x})= 0$, $i\in\mathcal I$, we consider  the utilitarian social welfare and analyze the classification of benefit or degradation of each agent $i$ based on $\Delta_i$.
For sufficiently small  interaction intensity $\phi$, the following results show that  this classification for agent $i$ can be determined by the amount of its   neighbors and the neighbors of   neighbors.

 \begin{figure}[t!]
\centering
\includegraphics[width=0.8\linewidth]{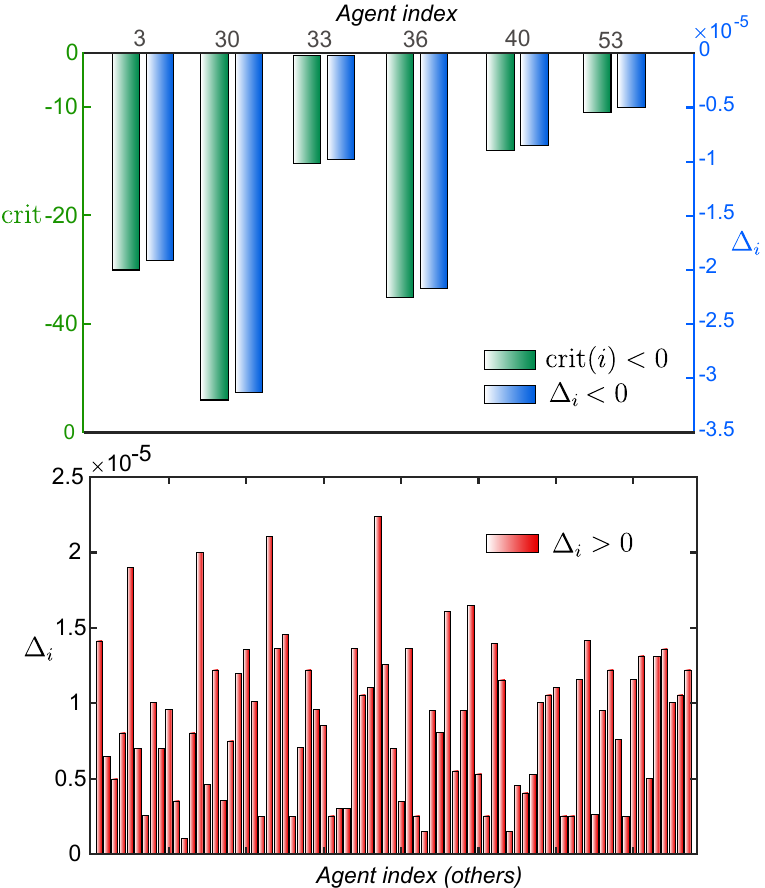}
\caption{
 The comparison of the criterion ${\rm crit}(i)$ (top) and  the actual $\Delta_i=f_i(\hat{\bf x})-f_i(x_{\rm bs})$ with  interaction intensity $\phi=0.001$ (bottom). 
}
\label{san_net}\vspace{-10pt}
\end{figure}

  \begin{thm}\label{pro1}Consider the networked game \eqref{netws} with utilitarian social welfare function $\mathcal W(x)=\mathcal W_{\rm u}(x)$. {There exists $\varrho>0$ such that  for any $\phi<\varrho$,} if 
\[
{\rm crit}(i):=2\sum\nolimits_k\sum\nolimits_l A_{il}A_{lk}-\big(\sum\nolimits_k A_{ik}\big)^2> 0 \ ({\rm resp}., <0).
\]
then    $\Delta_i=f_i(\hat{\bf x})-f_i(x_{\rm bs})>0$ ({\rm resp}., $<0$).
  \end{thm}
  \noindent\textbf{Proof}
See the Appendix below.$\hfill\square$

 \begin{figure*}[t!]
\centering
\includegraphics[width=0.85\linewidth]{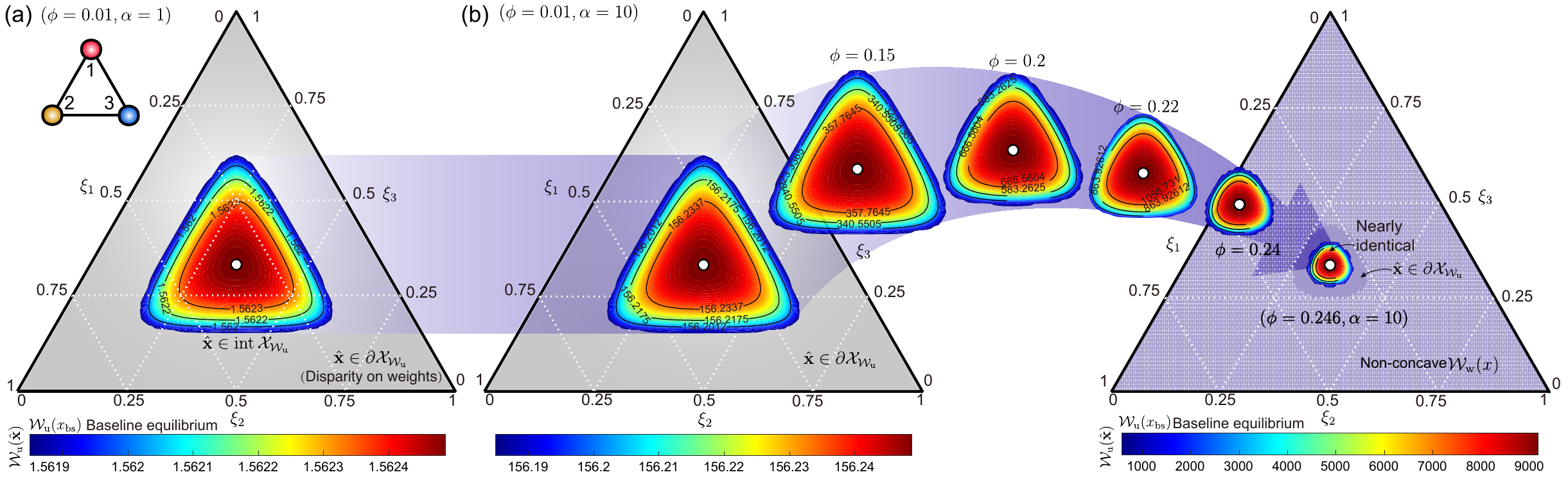}
\caption{{Ternary phase diagrams for a  three-agent R\&D networked game of the utilitarian social welfare $\mathcal W_{\rm u}(\hat{\bf x})$  with   $\mathcal W(x)=\mathcal W_{\rm w}(x)$}. 
}
\label{qqs1}\vspace{-10pt}
\end{figure*}

Interestingly, one can observe that    nodes \textcolor[rgb]{0,0,0}{with less neighbors of neighbors} are more likely to face degradation. For example,  the  firms  3, 30, 33, 36, 40, and 53   illustrated  as the red central nodes in Fig.~\ref{san_net1} are identified to satisfy ${\rm crit}(i)<0$ (see the verification  for $\phi=0.001$   in Fig.~\ref{san_net}).
In contrast, the improvement of the remaining nodes is sufficient to compensate for the degradation.
For example, for~regular graphs and three-node networks, the above criterion is always satisfied, implying that individual improvements  can be achieved without any redistribution at the steady state (i.e., optimal target state).

\subsubsection*{Discussion on influence of weights}We  illustrate   the agent priority paradox {with a  three-agent networked game} to indicate that the   phenomenon shown in Fig.~\ref{sasaasa} is not a special case.
 Fig.~\ref{qqs1} captures   ternary phase diagrams of the utilitarian social welfare $\mathcal W_{\rm u}(\hat{\bf x}(\xi))$. We see that  to guarantee  $\hat{\bf x}\in{\rm int}\, \mathcal X_{\mathcal W_{\rm u}}$ (or, equivalently, ${\mathcal W_{\rm u}}(\hat{\bf x})-{\mathcal W_{\rm u}}(x_{\rm bs})>0$), the welfare weights  $\xi$ of the social welfare function \eqref{weijt_f} cannot be too diverse. Otherwise,   $\hat{\bf x}\in\partial \mathcal X_{\mathcal W_{\rm u}}$ holds and hence a strict improvement over the baseline equilibrium $x_{\rm bs}$  is  impossible    for each agent.
Furthermore, by comparing the diagrams in  Figs.~\ref{qqs1}(a) and \ref{qqs1}(b), it is  observed  that  the region of weights $\xi$ with  a strict gain in utilitarian welfare remains essentially independent~of the marginal benefit $\alpha$.
But this region becomes visibly narrower as the interaction intensity $\phi$ increases  as in Fig.~\ref{qqs1}(b).
Therefore, stronger coupling among agents necessitates more uniform welfare weights  $\xi$  to gain   a strict improvement over the baseline equilibrium.
  
\subsubsection{Application to     Oligopoly Markets With Bertrand Competition and Price Regulation}\label{sec:ex2}

Consider a differentiated oligopoly market  composed of  $n$  firms producing  \emph{differentiated} products under  Bertrand competition  with the     payoff  
functions  \cite{haraguchi2016cournot}:
\begin{equation}
f_i(x)=(x_i-c_i)\tfrac{p_0(1-\delta)-(1+\delta(n-1))x_i+\delta\!\sum_{j\not=i}x_j}{\phi(1-\delta)(1+n\delta)},\  i\in\mathcal I,
\end{equation}
where     $x_i$, $c_i$, $p_0$,   $\phi$, and $\delta\in[0,1)$ are all positive parameters and
represent  the   price and the production cost  of     product     $i$,  the cap price,    
the market power, and  the degree of
product substitutability, respectively. 
Note that  lager $\delta\in[0,1)$ indicates  smaller differentiation among the products, i.e., when $\delta=0$, the  firms are selling   totally different types of products. 
The pseudo-gradient function   $F(x)=Rx+b$ 
is globally strongly monotone and
 the unique Nash equilibrium $x^*$ of $\mathcal G(f)$
 is  given by $x^*=-R^{-1} b$, where  $[R]_{ij}=\frac{\delta}{\phi(1-\delta)(1+n\delta)}>0$, $j\not=i$,  $[R]_{ii}=-\frac{2+2\delta(n-1)}{\phi(1-\delta)(1+n\delta)}<0$, $i\in\mathcal I$, and $ [b]_i=\frac{p_0(1-\delta)+c_i(1+\delta(n-1))}{\phi(1-\delta)(1+n\delta)}$, $i\in\mathcal I$. 

Now, let $n=4,\phi=2,p_0=6.8,c_1=1.6,c_2=1.8,c_3=2.3,c_4=3.2,\delta=0.2$ and consider the  social welfare function   $\mathcal W(x)=\mathcal W_{\rm w}(x)$
with  weights $\xi=[1,10,13,4]^\mathsf T/28$.  Take the baseline equilibrium $x_{\rm bs}$ to be      the Nash equilibrium $x^*$ without regulation, i.e.,  $x_{\rm bs}=x^*=[
3.2597,
    3.3538,
    3.5891,
    4.0127]^{\mathsf T}$.   In this case, note that     {Assumptions}~\ref{as_Ji}--\ref{as_social} and \ref{as_social0} hold  
with    $  \hat x^{\rm opt}\not\in \textcolor[rgb]{0,0,0}{\mathcal X_{\mathcal W_{\rm u}}}$ due to $ \mathcal W_{\rm u}(  \hat x^{\rm opt})=-0.2322< \mathcal W_{\rm u}( x_{\rm bs})=4.1619$.  
Furthermore,  the ideal optimal target state is obtained from  \eqref{hatx} as    $\hat{\bf x}=[4.2301,
    3.6985,
    3.8924,
    4.6479
]^\mathsf T$.

Next, since  \eqref{condition_thm} holds for $w=\xi$, $\hat \beta={\mathds 1}_4$,  $K={\bf 0}_{n\times n}$ with  a minimal eigenvalue $\zeta$  of $\frac{R+R^{\mathsf T}}{2}$,  it follows from Theorem~\ref{smalX} that the  linear   incentive mechanism \eqref{netw}, \eqref{incen_infla} with the update rules  \eqref{xrm}, \eqref{ut}, \eqref{ztt},  \eqref{ysinkal2}, and $K_i={\bf 0}_{n\times n}$  guarantees  that the state  $x(t)$ of agents   under the pseudo-gradient dynamics \eqref{dynamics2} globally  converges to the  {optimal target state}  $\hat{\bf x}$ with balanced budget and individual rationality   for any  sensitivity parameters $\beta_i\in\mathbb R_+$, $i\in\mathcal I$.
In the simulation, let  $\epsilon(0)={\mathds 1}_n$ and  $\lambda(0)=0$. 
Fig.~\ref{ex2_1} illustrates  the trajectories of the state $x(t)$, the effective  payoff values $V_i(x(t))$,  the incentive amounts $y_i(x(t))$, and the social welfare $\mathcal W(x(t))$  with $\eta_i=\xi_i$, $i\in\mathcal I$, under 10 sets of different  sensitivity parameters $\beta_i\in[0.6,5.6]$, $i\in\mathcal I$,   and  random initial state $x(0)=x_{\rm r}(0)\in[3,5.2]^4$.
 It can be seen   that all the trajectories of the state $x(t)$ converge to  $\hat{\bf x}$ with balanced budget, while the effective payoffs $V_i(x(t))$ converge  to the same level as the baseline equilibrium  without incentives. This implies that the   mechanism is individually rational at $t\to\infty$  and thereby verifies  the results of  Theorem~\ref{smalX}. 
Moreover, the proposed mechanism substantially  improves the social welfare $\mathcal W(x)$ at the equilibrium outcome  in Fig.~\ref{ex2_1}(d).
It is also interesting to notice that 
 the commodity prices of     agents  at $\hat{\bf x}$ are increased compared to the baseline equilibrium $x_{\rm bs}$, i.e., $\hat{\bf x}_i>x_i^*$, $i\in\mathcal I$. 
Therefore, the proposed mechanism  achieves  a price control of commodities  without harming the payoffs of   companies, which can be potentially used for  inflation/deflation control in economics.

    \begin{figure}[t!]
\begin{center}
		\includegraphics[width=7cm]{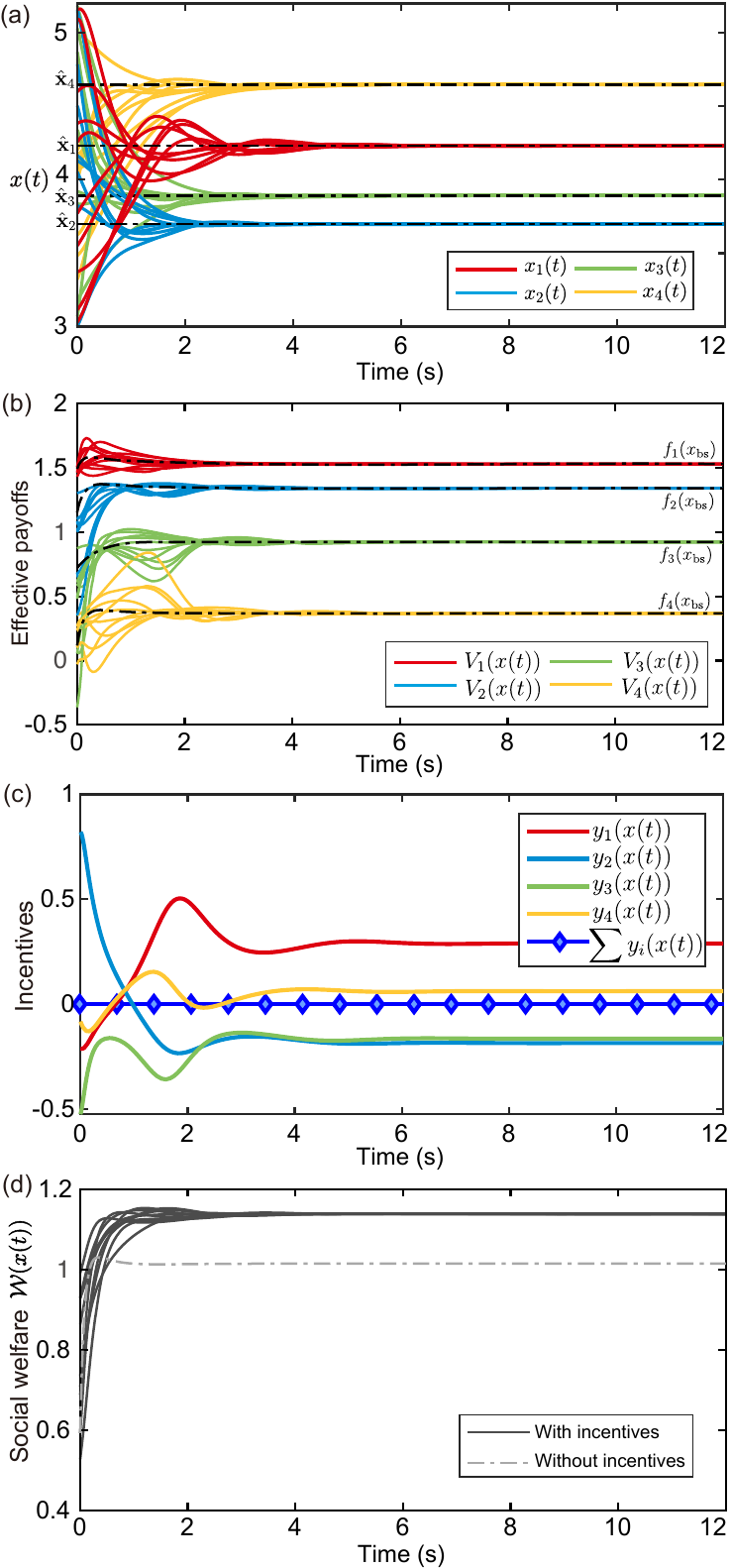}
	\caption{The performance of the proposed incentive mechanism for a four-agent Bertrand  oligopoly game under     the weighted social welfare function. (a) (resp., ({b})) Trajectories of the states $x(t)$ (resp., payoffs)  with and without the incentive mechanism.     (c)  Trajectories of incentives. (d)   Trajectories  of the social welfare $\mathcal W(x(t))$ with and without the incentive mechanism.  The dash-dotted curves in  (b) represent the trajectories of        payoffs $f_i(x(t))$ under pseudo-gradient dynamics of the un-incentivized system. }
\label{ex2_1}\vspace{-25pt}
\end{center}
\end{figure}

\subsubsection{Application to   Oligopoly Markets   With Cournot  Competition and Pollution Reduction}
Consider a similar differentiated oligopoly market composed of  $n$  firms producing and selling  \emph{differentiated} products with the product substitutability    
 $\delta\in[0,1)$.     In this part,  we study  Cournot   competition    in which   quantities of products  are the decision variables. 
It is assumed that  each firm $i$ needs $\tau x_i$  amount of electricity to produce      $x_i$ quantity of products  with the energy purchase cost $c_i\in\mathbb R_+$, $i\in\mathcal I$.
 The payoff functions  are 
given from   \cite{haraguchi2016cournot} by
\begin{equation}
f_i(x)=x_i(-\tau c_i+\underbrace{p_0-\phi x_i-\sum\nolimits_{j\not=i}\phi\delta x_j}_{ p_i(x)}),\  i\in\mathcal I,
\end{equation}
where      $p_0\in\mathbb R_+$ is the
  cap price,  $\phi\in\mathbb R_+$  is the     market power, and the  term   $p_i(x)$ is called  the   market   price  function.
Note that the market price function $p_i(x)$ reflects the fact that  the  prices tend to be mutually independent  when the firms  sell more differentiated products (with a smaller $\delta$). 
The pseudo-gradient function $F(x)$  is given  by the affine function $F(x)=R x+b$ with  $[R]_{ij}=-\phi\delta<0$, $j\not=i$,  $[R]_{ii}=-2\phi<0$,   $i\in\mathcal I$, and $ b_i=p_0-\tau c_i$, $i\in\mathcal I$. Note that this game is   strongly monotone   since  the eigenvalues  of the symmetric Jacobian matrix $R$ are all negative as  $\lambda_1=\cdots=\lambda_{n-1}=\phi(\delta-2)$ and $\lambda_n=-\beta(2+(n-1)\delta)$ for  any   $\phi\in\mathbb R_+$,  $\delta\in[0,1)$.
Thus,  the Nash equilibrium without regulation  is uniquely    given by $x_{\rm bs}=-R^{-1} b$.

Now,  let $n=4,\tau=1,p_0=9.1, g=0.2,\delta=0.2,c_1=1.6,c_2=1.8,c_3=2.3,c_4=3.2,\phi=2$, and consider the  social welfare function   $\mathcal W(x)=\mathcal W_{\rm w}(x)$
with the normalized  weights $\xi=[1,1,2,3]^\mathsf T/7$. Note that   firms 3 and 4 use  \emph{clean energy} rather than traditional energy and thus  \textcolor[rgb]{0,0,0}{incur  higher} costs than firms 1 and 2. 
Let the baseline equilibrium $x_{\rm bs}$ be      the Nash equilibrium $x^*$ without regulation, i.e.,  $x_{\rm bs}=x^*=[
    1.4957,
    1.4402,
    1.3013,
    1.0513]^{\mathsf T}$.   In this case, note that     {Assumptions}~\ref{as_Ji}--\ref{as_social} and \ref{as_social0} hold.   
To reduce  the total pollution ${\mathds 1}_n^\mathsf T x$ caused by production, we construct the environmental social welfare function \cite{shan2025environmental} as 
\begin{equation}\label{sqml_sw}
\mathcal W(x)=\mathcal W_{\rm w}(x)-g{\mathds 1}_n^\mathsf T x,
\end{equation}
which  balances aggregate industry profits against environmental damage from total production with  $g>0$ representing  an environmental coefficient.

\begin{figure}[t!]
\centering
\includegraphics[width=0.99\linewidth]{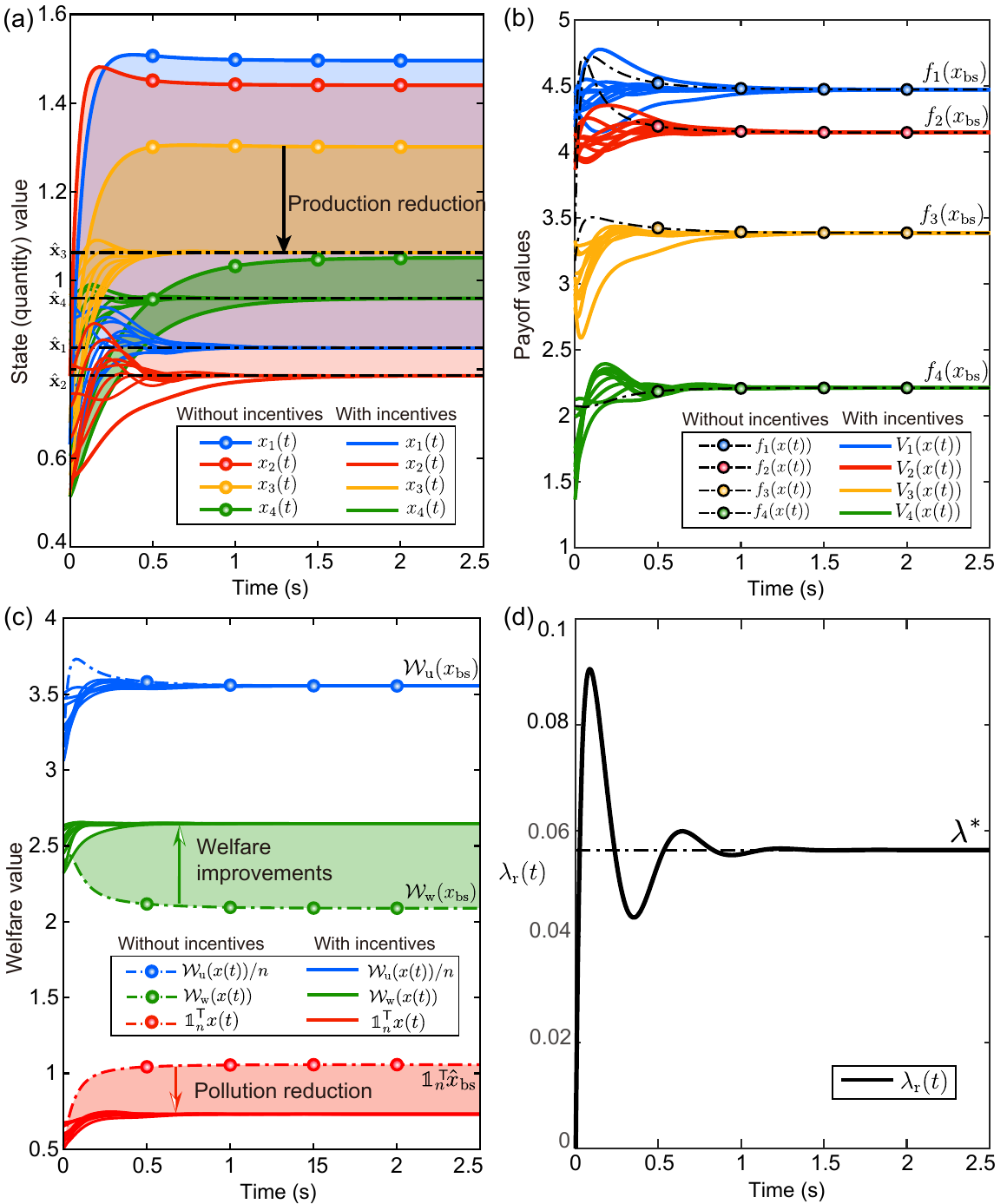}
\caption{{The performance of the proposed  incentive mechanism    for a  four-agent Cournot oligopoly  game  under the environmental social welfare function \eqref{sqml_sw}}.  
(a) (resp., ({b})) Trajectories of the states $x(t)$ (resp., payoffs)  with and without the incentive mechanism.
(c) Trajectories of the social welfare $\mathcal W(x(t))$, the averaged  welfare $\mathcal W_{\rm u}(x(t))/n$ and total  pollution $\mathds{1}_n^\mathsf Tx(t)$ with and without the  incentive mechanism.
(d) Trajectory of the Lagrange multiplier  $\lambda_{\rm r}(t)$ in reference system \eqref{xrm}.
}
\label{diff_mark}\vspace{-10pt}
\end{figure}

Next, since \eqref{condition12} holds for $\hat \beta={\mathds 1}_4$ with  a minimal eigenvalue $\zeta$  of $\frac{R+R^{\mathsf T}}{2}$,  it follows from  Corollary~\ref{cor11} that the  linear incentive functions  \eqref{incen_infla}   with the updating  rules   \eqref{xrm}, \eqref{ut},   \eqref{ysinkal2},   and \eqref{ztt2}   guarantee that the state  $x(t)$ of agents   under the pseudo-gradient dynamics \eqref{dynamics2} globally  converges to the   {optimal target state}  $\hat{\bf x}$ with balanced budget and individual rationality     for any  sensitivity parameters $\beta_i\in\mathbb R_+$, $i\in\mathcal I$.
Let  $\epsilon(0)={\mathds 1}_n$ and  $\lambda(0)=0$. 
Fig.~\ref{diff_mark} shows   the trajectories of the state $x(t)$, the effective  payoff values $V_i(x(t))$,  the social welfare $\mathcal W(x(t))$, and the total pollution  $\mathds{1}_nx^\mathsf T(t)$, whereas  the  incentive amounts $y_i(x(t))$ are shown in Fig.~\ref{diff_mark2}(a), with $\eta_i=\xi_i$, $i\in\mathcal I$, under 10 sets of different  sensitivity parameters $\beta_i\in[0.6,5.6]$, $i\in\mathcal I$,   and random initial states $x(0)=x_{\rm r}(0)=\in[0.5,1]^4$.  It can be seen from these figures that all the trajectories of the state $x(t)$ converge  to  $\hat{\bf x}$ with balanced budget, while the incentivized payoff values  $V_i(x(t))$ converge to the same level as the one without incentives, which implies that the   mechanism is individually rational  at $t\to\infty$ and thereby verifying the results of  Corollary~\ref{cor11}. 
Moreover, by substantially  improving  the social welfare $\mathcal W(x)$, the proposed mechanism suppresses    the  quantity of  products for each firm, thereby reducing the total pollution $\mathds{1}_nx^\mathsf T(t)$ at the equilibrium outcome.

\begin{figure}[t!]
\centering
\includegraphics[width=0.6\linewidth]{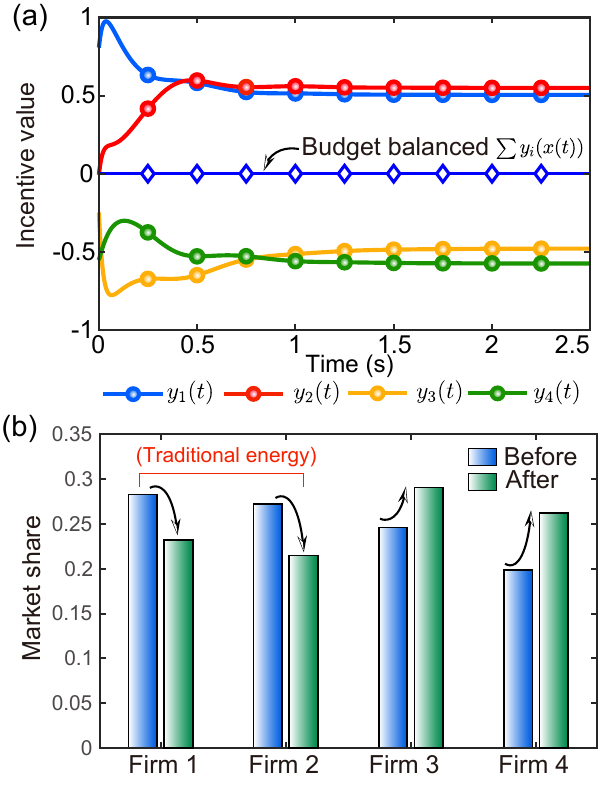}
\caption{Trajectories of incentive values (a) and the steady state market shares (b)  of the four companies before  and after incentive mechanisms. 
Note that the market share of firm $i\in\mathcal I$ is calculated by $[{x_{\rm bs}}]_i/(\mathds 1_n^\mathsf T x_{\rm bs})$ or $\hat{\bf x}_i/(\mathds 1_n^\mathsf T\hat{\bf x})$ in (b). 
}
\label{diff_mark2}\vspace{-10pt}
\end{figure}

\begin{figure*}[t!]
\begin{center}
\includegraphics[width=0.85\linewidth]{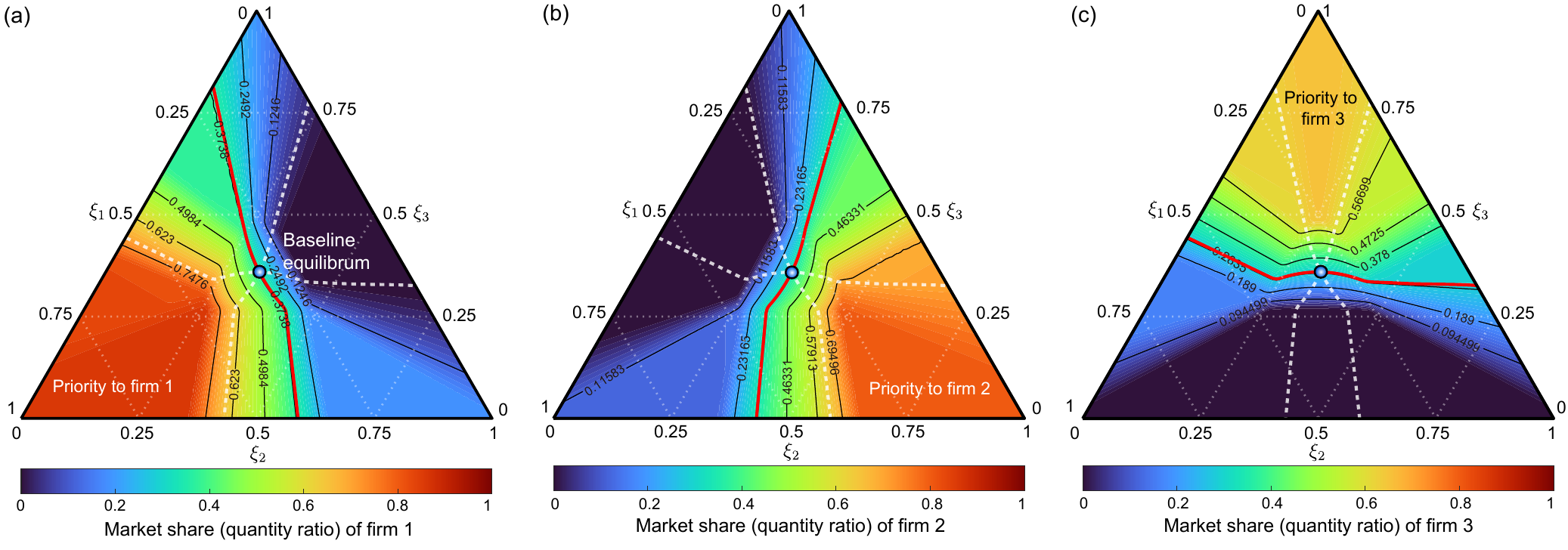}
\captionsetup{justification=justified}
\caption{Ternary phase diagrams of the market shares at the optimal target state $\hat{\bf x}$  versus welfare weights of a  three-agent Cournot oligopoly  game. The red curves in (a), (b), and (c) represent the welfare weights $\xi$ where the market shares of firms 1, 2, and 3 at  $\hat{\bf x}$    are equivalent to that of     baseline equilibrium $x_{\rm bs}$, respectively. The   blue node denotes the  welfare weight  $\xi$ generating   the exactly same market shares  of all the firms   at  $\hat{\bf x}$   as  the baseline equilibrium $x_{\rm bs}$.}\label{diff_3phase}
\end{center}\vspace{-18pt}
\end{figure*}

\subsubsection*{Discussion on influence of weights}It is  interesting to see from Figs.~\ref{diff_mark}(a) and \ref{diff_mark2}(b) that the extent of quantity reduction  appears to be closely related to the welfare weights $\xi$ assigned to the agents in the environmental social welfare function \eqref{sqml_sw}.
For example, the firm with the largest weight (firm~4) exhibits the smallest reduction in production quantity.
It can be further seen from the market shares displayed in Fig.~\ref{diff_mark}(b) that firms 3 and 4, who use clean energy and hence are given double and triple priorities, respectively, gain   market shares from firms 1 and 2 without incurring  payoff  loss to firms 1 and 2.

This fact is   further investigated in  Fig.~\ref{diff_3phase},  which visualizes the influence of welfare  weights $\xi$ in \eqref{sqml_sw} on the market shares for a 3-firm market with $\tau=1,\phi=2, p_0=9.1, g=0.2,\delta=0.6, c_1=1.6,c_2=1.8,c_3=2.3$. 
It can be seen from this  ternary phase diagram that given a larger weight $\xi_i$ for any firm $i$ could enlarge  its  market share $\hat{\bf x}_i/(\mathds 1_n^\mathsf T\hat{\bf x})$ at the regulated steady state $\hat{\bf x}$.
We also observe that the \textcolor[rgb]{0,0,0}{largest} market shares depends on the production cost $c_i$ (see different largest market share of the firms with cost in Figs.~\ref{diff_3phase}(a)--(c)). The  \textcolor[rgb]{0,0,0}{largest} market share of the firm with cost advantages (e.g., firm 1~who possesses the lowest production cost) is higher than  other firms.

\vspace{-15pt}
\section{Conclusion}\label{section6}

We investigated the incentive design problem for noncooperative dynamical systems with an individually rational  incentive   mechanism. 
In the proposed incentive mechanism, the system regulator uses zero-sum incentive functions to shift the Nash equilibrium of the game and to guarantee   Pareto efficiency at the target state which maximizes the social welfare function under the restrictions of     budget balance and individual rationality.
A constrained optimization problem was considered to determine the optimal  target  state of IRPE incentives. For  general social welfare functions, we revealed  that  the   utilitarian social welfare function is always involved to define the constraint. 
Our analytical results demonstrated that simple {linear-quadratic} forms are sufficient as incentive function candidates.
Sufficient conditions of the updating laws of the incentive functions were presented to  ensure that the state of agents converges to the optimal target state as a Pareto-efficient Nash equilibrium with individual rationality and balanced budget. 
The applications of our results in networked games and differentiated oligopoly markets were considered.
We numerically observed that our results contribute to  reshaping the  market configurations (e.g.,  deflation control, market share control) without harming the payoffs of companies.


  \vspace{-8pt}
\bibliographystyle{IEEEtran}        
\bibliography{autosam}

  \appendix[Proof of Theorem~\ref{pro1}]
 It follows using the Neumann series  that $x_{\rm bs}=\alpha(I-\phi A)^{-1} 1_n=\alpha\sum_{p=0}^\infty \phi^p A^p 1_n$ and  the socially optimal state  is    $\hat x^{\rm opt}=\alpha(I-2\phi A)^{-1}1_n=\alpha\sum_{p=0}^\infty 2^p\phi^p A^p1_n$, where  $(i,j)$-entry of $A^p$ indicates  the number of walks of length $p$ between nodes $i$ and $j$.
From the first order conditions,    the   payoffs of  $ f_i(x_{\rm bs})$ and $f_i(\hat{\bf x}) $  are respectively given by 
\begin{align}\nonumber
f_i(x_{\rm bs})&=-\tfrac{1}{2}(x_{\rm bs})_i^2+x_i\Big(\phi\sum_{j\not=i} [A]_{ij}(x_{\rm bs})_j+\alpha\Big)=\frac{1}{2}(x_{\rm bs})_i^2,
\\ \nonumber
f_i(\hat{\bf x})&=-\tfrac{1}{2}(\hat{\bf x}_i)^2+(\hat{\bf x}_i)\Big(\phi\sum\nolimits_{j\not=i} [A]_{ij}(\hat{\bf x})_j+\alpha\Big)\\ \nonumber
&=-\tfrac{1}{2}(\hat{\bf x}_i)^2+(\hat{\bf x}_i)\Big(\phi (A\hat{\bf x})_i+(\hat{\bf x}_i)-2\phi (Ax)_i\Big)\\ \nonumber
&=\tfrac{1}{2}(\hat{\bf x}_i)^2-\phi(\hat{\bf x}_i)(A\hat{\bf x})_i.
\end{align}
Using $\hat{\bf s}=\frac{1}{\alpha}\hat{\bf x}$ and $s_{\rm bs}=\frac{1}{\alpha}x_{\rm bs}$, we have  that 
$
\Delta_i=\frac{1}{2}[(\hat{\bf x}_i)^2-(x_{\rm bs})_i^2]-\phi(\hat{\bf x}_i)(A\hat{\bf x})_i
=\frac{1}{2}\alpha^2[(\hat{\bf s})_i^2-(s_{\rm bs})_i^2]-\alpha^2\phi(\hat{\bf s})_i(A\hat{\bf s})_i
$.
Now, using expanded series
\begin{align}\nonumber
( s_{\rm bs})_i&=\left(\sum\nolimits_{p=0}^\infty \phi^p A^p1_n\right)_i,\ \ 
(\hat{\bf s})_i=\left(\sum\nolimits_{p=0}^\infty 2^p\phi^p A^p1_n\right)_i,\\ \nonumber
(A\hat{\bf s})_i&=\left(A\sum\nolimits_{p=0}^\infty 2^p\phi^p A^p1_n\right)_i\end{align}
we obtain 
 \begin{align}\nonumber
 [(\hat{\bf s})_i^2-(s_{\rm bs})_i^2]&=2\phi\sum\nolimits_k A_{ik}+3\phi^2\big(\sum\nolimits_k A_{ik}\big)^2\\ \nonumber
 &+6\phi^2\sum\nolimits_k\sum\nolimits_l A_{il}A_{lk}+O(\phi^3),\\ \nonumber
 \phi(\hat{\bf s})_i(A\hat{\bf s})_i&=\phi\sum\nolimits_k A_{ik}+2\phi^2\sum\nolimits_k\sum\nolimits_l A_{il}A_{lk}\\ \nonumber
 &+2\phi^2\big(\sum\nolimits_k A_{ik}\big)^2+O(\phi^3),
 \end{align}
with $O(\phi^3)$ representing the higher order terms, which imply 
\begin{align} 
\Delta_i
=\alpha^2 \Bigg[\frac{1}{2}{\rm crit}(i)\phi^2+O(\phi^3)\Bigg].
\end{align}
Therefore, when $\phi$ is infinitesimal, the sign of $\Delta_i$ is strictly positive  (resp., negative) if ${\rm crit}(i)>0$ (resp., ${\rm crit}(i)<0$), which completes the proof.  $\hfill\square$

\begin{IEEEbiography}[{\includegraphics[width=1in,height=1.25in,clip,keepaspectratio]{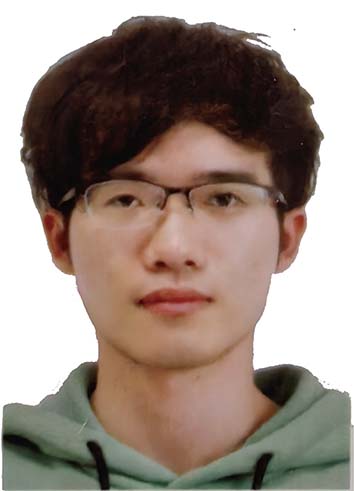}}]{Yuyue Yan}
(S'19--M'22) received the B.Eng. degree in electronic information engineering from Xiamen University Tan Kah Kee College, Fujian,  China, in 2017
 the M.E. degree and the Ph.D. degree in systems and control engineering from Tokyo Institute of Technology, Tokyo, Japan, in 2019 and 2022, respectively.
He was a  Research Fellow with the Department of Systems and Control Engineering, Tokyo Institute of Technology, Japan in 2022 and 2023, and  with the Graduate School of Engineering, Osaka University, Osaka, Japan in 2024 and 2025. 
He is currently a Specially Appointed Assistant Professor with the Department of Information Physics and Computing, the University of Tokyo, Tokyo, Japan.

He is a member of IEEE IES Technical Committee on Network-Based Control Systems and Applications. 
His research interests include stability of nonlinear dynamical systems, hybrid systems, and noncooperative system; and applications to multiagent systems, atomic clock systems, financial dynamics, and economic/social systems.
\end{IEEEbiography}

\begin{IEEEbiography}[{\includegraphics[width=1.1in,height=1.25in,clip,keepaspectratio]{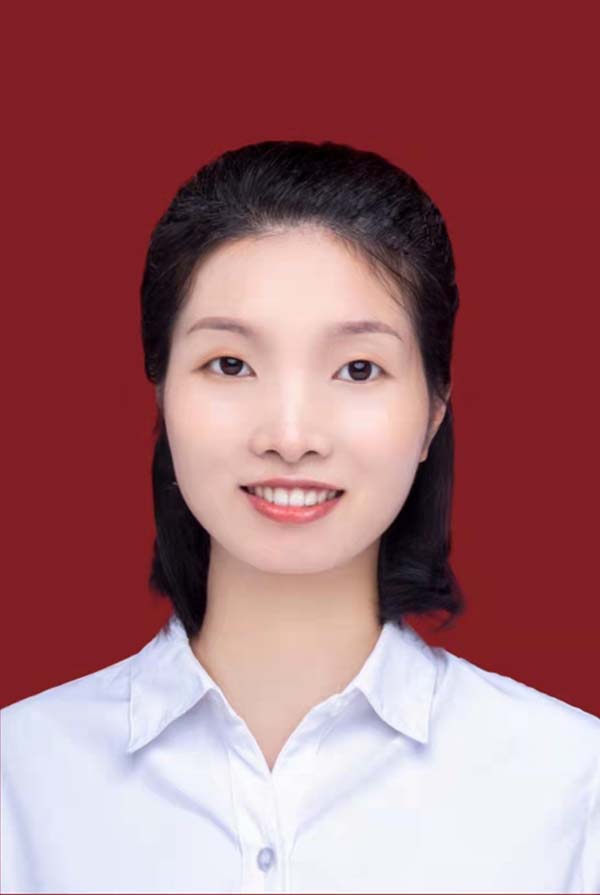}}]{Maojiao Ye}
 (M'17-SM'23) received the B.Eng. degree in Automation from the University of Electronic Science and Technology of China, Sichuan, China, in 2012 and the Ph.D. degree from Nanyang Technological University, Singapore, in 2016. 
 She is currently a Professor with the School of Automation, Nanjing University of Science and Technology. Prior to her current position, she was a Research Fellow in the School of Electrical and Electronic Engineering at Nanyang Technological University from 2016-2017. 
Her research interests include game theory, distributed optimization, and their applications.  

Professor Ye was a recipient of the Young Scientist Award from the Chinese Association of Automation in 2023, Guan Zhao-Zhi Best Paper Award in the 36th Chinese Control Conference 2017, and the Best Paper Award in the 15th IEEE International  Conference on Control and Automation 2019.  Professor Ye is an Associate Editor  of IEEE Transactions on Industrial Informatics, Control Engineering Practice, and IEEE/CAA Journal of Automatica Sinica. She is the Vice-Chair of IEEE IES Technical Committee on Network-Based Control Systems and Applications, and a member of IEEE CSS Conference Editorial Board.
\end{IEEEbiography}

\begin{IEEEbiography}[{\includegraphics[width=1in,height=1.25in,clip,keepaspectratio]{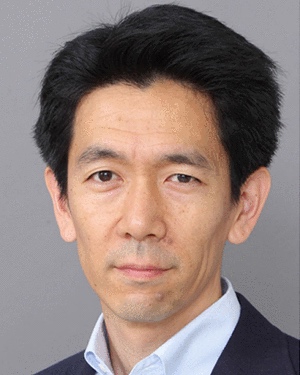}}]{Hideaki Ishii}
(M’02-SM’12-F’21) received the
M.Eng. degree in applied systems science from Kyoto University, Kyoto, Japan, in 1998, and the Ph.D. degree in electrical and computer engineering from the University of Toronto, Toronto, ON, Canada, in 2002. He was a Postdoctoral Research Associate at the University of Illinois at Urbana-Champaign, Urbana, IL, USA, from 2001 to 2004, and a Research Associate at The
University of Tokyo, Tokyo, Japan, from 2004 to 2007. He was an Associate Professor and Professor at the Department of Computer Science, Tokyo Institute of Technology, Yokohama, Japan in 2007–2024. Currently, he is a Professor at the Department of Information Physics and Computing at The University of Tokyo, Tokyo, Japan. He was a Humboldt Research Fellow at the University of Stuttgart in 2014–2015. He has also held visiting positions at CNR-IEIIT at the Politecnico di Torino, the Technical University of Berlin, and the City University of Hong Kong. His research interests include networked control systems, multiagent systems, distributed algorithms, and cyber-security of control systems.

Dr. Ishii has served as an Associate Editor for Automatica, the IEEE Control Systems Letters, the IEEE Transactions on Automatic Control, the IEEE Transactions on Control of Network Systems, and the Mathematics of Control, Signals, and Systems. He was a Vice President for the IEEE Control Systems Society (CSS) in 2022–2023. He was the Chair of the IFAC Coordinating Committee on Systems and Signals in 2017–2023. He served as the IPC Chair for the IFAC World Congress 2023 held in Yokohama, Japan. He received the IEEE Control Systems Magazine Outstanding Paper Award in 2015. Dr. Ishii is a  Fellow of IEEE and IFAC.
\end{IEEEbiography}

\begin{IEEEbiography}[{\includegraphics[width=1in,height=1.25in,clip,keepaspectratio]{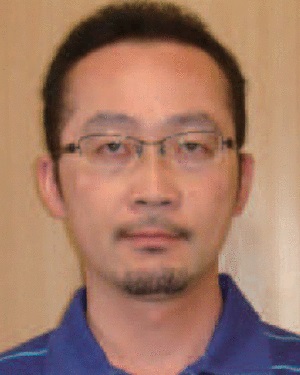}}]{Tomohisa Hayakawa}
(S'00--M'04)
received the B.Eng. degree in aeronautical engineering from Kyoto University, Kyoto, Japan, in 1997, the M.S. degree in aerospace engineering from The State University of New York, Buffalo, NY, USA, in 1999, the M.S. degree in applied mathematics and the Ph.D. degree in aerospace engineering from the Georgia Institute of Technology, Atlanta, GA, USA, in 2001 and 2003, respectively.

He was a Research Fellow with the Department of Aeronautics and Astronautics, Graduate School of Engineering, Kyoto University and with the Japan Science and Technology Agency in 2006, and then joined the Tokyo Institute of Technology, Tokyo, Japan, where he is currently an Associate Professor. His research interests include stability of nonlinear systems; noncooperative systems; hybrid systems; nonlinear adaptive control; neural networks and intelligent control; stochastic dynamical systems; and applications to aerospace vehicles, formation control of multi-agent systems, financial dynamics, and economic/social systems in a future smart society.
\end{IEEEbiography}

\end{document}